\documentclass[12pt, draftclsnofoot, onecolumn]{IEEEtran}
\usepackage[T1]{fontenc}
\usepackage{graphicx}
\usepackage{amsmath, amssymb, ntheorem, pstricks, mathrsfs, subfig}
 \usepackage{pst-node}
 \usepackage{pst-blur}
 \usepackage[font=small]{caption}
\usepackage{algorithm}
\usepackage{color}
\usepackage{caption}
\usepackage[noend]{algpseudocode}
\usepackage[normalem]{ulem}
\usepackage{cite}

\newtheorem{theorem}{Theorem}

\newtheorem{proposition}{Proposition}

\allowdisplaybreaks

\definecolor{mygreen}{rgb}{0, 0.5, 0}

\definecolor{superlightgray}{rgb}{.9,0.9,0.9}

\newcommand{\vechi}{\protect\overrightarrow{\vspace{5.8pt}}\hspace{-9pt}\mathbf{h}_i}
\newcommand{\vech}{\protect\overrightarrow{\vspace{5.8pt}}\!\!\!\!\!\mathbf{h}}

\title{Interleaving Channel Estimation and Limited Feedback for Point-to-Point Systems with a Large Number of Transmit Antennas}
\author{Erdem Koyuncu, Xun Zou, and Hamid Jafarkhani
\thanks{
This work was supported in part by the NSF Award ECCS-1611575.}\thanks{ This work was presented in part at the IEEE International Symposium on Information Theory (ISIT) in July 2015 \cite{confversion}.}\thanks{
E. Koyuncu is with Department of Electrical and Computer Engineering, University of Illinois at Chicago (e-mail: ekoyuncu@uic.edu).
X. Zou and H. Jafarkhani are with Center for Pervasive Communications \& Computing,  University of California, Irvine (e-mails: xzou4@uci.edu; hamidj@uci.edu).
}}
\begin{document}
\maketitle
\vspace{-50pt}
\begin{abstract}
We introduce and investigate the opportunities of multi-antenna communication schemes whose training and feedback stages are interleaved and mutually interacting. Specifically, unlike the traditional schemes where the transmitter first trains all of its antennas at once and then receives a single feedback message, we consider a scenario where the transmitter instead trains its antennas one by one and receives feedback information immediately after training each one of its antennas. The feedback message may ask the transmitter to train another antenna; or, it may terminate the feedback/training phase and provide the quantized codeword (e.g., a beamforming vector) to be utilized for data transmission. As a specific application, we consider a multiple-input single-output system with $t$ transmit antennas, a short-term power constraint $P$, and target data rate $\rho$. We show that for any $t$, the same outage probability as a system with perfect transmitter and receiver channel state information can be achieved with a feedback rate of $R_1$ bits per channel state and via training $R_2$ transmit antennas on average, where $R_1$ and $R_2$ \textit{are independent of $t$}, and depend only on $\rho$ and $P$. In addition, we design variable-rate quantizers for channel coefficients to further minimize the feedback rate of our scheme.

\end{abstract}

\vspace{-2pt}

\section{Introduction}\vspace{-2pt}
The performance of a wireless communication system can be greatly improved by making the channel state information (CSI) available at the transmitter and the receiver. In a massive multiple-input single-output (MISO) system, having CSI at the transmitter (CSIT) is especially desirable as one can then fully exploit the performance gains promised by the large number of transmit antennas via CSI-adaptive transmission strategies such as beamforming. A typical way to acquire CSIT is channel estimation followed by (digital) feedback.

\begin{figure}[h]
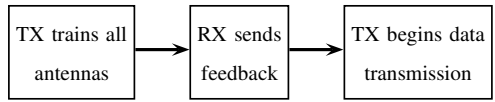

\centering
\scalebox{0.8}{
\psset{shadowcolor=black!70,blur=true}
\hspace{18pt}\begin{psmatrix}[rowsep=0.5,colsep=0.9]
\psframebox{\small \begin{tabular}{{@{}c@{}}} TX trains all \\ antennas \end{tabular}} &
\psframebox{\small\begin{tabular}{{@{}c@{}}} RX sends \\ feedback \end{tabular}} &
\psframebox{\small\begin{tabular}{{@{}c@{}}} TX begins data \\ transmission \end{tabular}} &
\psset{arrows=->,fillstyle=none,linewidth = 1.5pt,arrowsize=6pt}\ncline{1,1}{1,2}\ncline{1,2}{1,3}
\end{psmatrix}}
\caption{Conventional training and limited feedback. TX and RX stand for the transmitter and the receiver, respectively.}
\label{figconventional}
\end{figure}

Channel training/estimation and feedback are traditionally viewed as two non-interleaving processes, as shown in Fig. \ref{figconventional}. According to this traditional viewpoint, for each channel state, the transmitter first trains all of its antennas at once, so that the receiver acquires the entire CSI (or, in general, an erroneous version thereof.). This initial training phase is followed by the receiver feeding back a possibly-quantized version of the CSI. The receiver's feedback is then utilized at the transmitter side for data transmission (e.g., as a quantized beamforming vector.).
Designing such limited feedback systems is a fundamental problem of communication theory and has been the subject of many publications \cite{love3}. In particular, limited feedback beamforming \cite{narula1} has been studied through several different approaches that utilize Grassmannian line packings \cite{love1}, vector quantization \cite{roh1}, combinations with orthogonal \cite{jongren1} or quasi-orthogonal \cite{LLHJ05} space-time codes, variable-length coding \cite{koyuncu1}, or other systematic constructions \cite{mukkavilli1}. Conditions to achieve full diversity in a finite feedback scheme has been discussed in \cite{rev1,rev2}. Various distributed limited feedback schemes \cite{dlf1,dlf5,dlf2,dlf3,dlf4} provide generalizations to multi-user networks.

The conventional scheme in Fig. \ref{figconventional} appears to be infeasible in the case of a massive MISO system. Even the channel training/estimation phase, by itself, would be very challenging to realize due to the large number of transmit antennas that need to be trained. Moreover, even if one assumes that the training stage somehow comes with no cost, feeding back the associated large number of channel values to the transmitter appears to be infeasible. Conventional limited feedback schemes also do not provide much hope in this context: The feedback rates required for even the simplest of the limited feedback schemes such as antenna selection grow without bound as the number of transmit antennas grows to infinity. In \cite{jaobHowmany}, it is analyzed in detail how many antennas per user terminal are needed to achieve some percentage of the ultimate performance limit with infinitely many antennas.

There has been some work on channel estimation and CSI feedback in massive MIMO systems; a survey can be found in \cite{survey}. In particular, \cite{junilNTCQ} proposes a noncoherent trellis-coded quantization scheme, whose encoding complexity scales linearly with the number of antennas. In \cite{RaoCS, kuoCS, anothercspaper}, compressive sensing techniques are utilized to reduce the feedback overhead of the CSI estimation. In addition, several studies \cite{junilTrainWithMemo, junyoungJoint, byungjuGroup,gaoSpatial} have demonstrated that channel or antenna correlation can be exploited to reduce the overhead of the downlink training phase. A multi-beam selection scheme for massive MIMO is presented in \cite{hybridmimo}. The problem of designing training sequences with low overheads have been studied in \cite{snoh1}.  There are also several other approaches proposed for resolving the challenges of training and limited feedback in the more general context of multi-user MIMO; see e.g., \cite{jaf1, kob1, salim1, kang1}.

\begin{figure}[h]
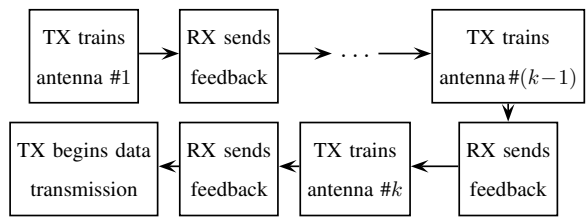

\centering
\scalebox{0.8}{
\psset{shadowcolor=black!70,blur=true}
\begin{psmatrix}[rowsep=0.3,colsep=0.34]
\psframebox{\small\begin{tabular}{{@{}c@{}}} TX trains \\ antenna \#$1$ \end{tabular}} &
\psframebox{\small\begin{tabular}{{@{}c@{}}} RX sends \\ feedback \end{tabular}} &
$\,\,\Huge{\cdots}\,\,$ &
\psframebox{\small\begin{tabular}{{@{}c@{}}} TX trains \\ antenna$\!$ \#$(k\!-\!1)$ \end{tabular}} \\
 \psframebox{\small\begin{tabular}{{@{}c@{}}} TX begins data \\ transmission \end{tabular}} &
 \psframebox{\small\begin{tabular}{{@{}c@{}}} RX sends \\ feedback \end{tabular}} &
 \psframebox{\small\begin{tabular}{{@{}c@{}}} TX trains \\ antenna \#$k$ \end{tabular}}  &
 \psframebox{\small\begin{tabular}{{@{}c@{}}} RX sends \\ feedback \end{tabular}} &
\psset{arrows=->,fillstyle=none,arrowsize=6pt}\ncline{1,1}{1,2}\ncline{1,2}{1,3}\ncline{1,3}{1,4}\ncline{1,4}{2,4}\ncline{2,4}{2,3}\ncline{2,3}{2,2}\ncline{2,2}{2,1}
\end{psmatrix}}
\caption{Interleaved training and limited feedback. The number of trained antennas $k$ varies from one channel state to another, and is itself decided through the training and feedback phases.}
\label{figinterleaved}
\end{figure}

Our proposed solution is to interleave the training and feedback stages as shown in Fig. \ref{figinterleaved}. Unlike the conventional scheme in Fig. \ref{figconventional},  the transmitter trains its antennas one by one and receives feedback information after training each one of its antennas. A feedback message may ask the transmitter to train another antenna (and also provide side information about the channel state), or it may result in the termination of the training phase, in which case it also provides the quantized codeword to be utilized by the transmitter for data transmission.

An interleaved scheme offers the following unique opportunity: If the already-trained antennas provide sufficiently favorable conditions for data transmission, one can then terminate the training phase and thus avoid wasting more resources on training the rest of the antennas. One main message of this paper is that in certain scenarios, we can make use of this opportunity to design multi-antenna communication systems whose feedback and training overheads remain completely independent of the number of transmit antennas, and which, at the same time, can achieve the same outage performance as a system with perfect transmitter and receiver CSI. Specifically, we consider here a single-user point-to-point MISO system with the outage probability performance measure. Note that, while the ``mainstream'' use case of a massive transmitter antenna array is to support multiple users, a single-user system suffering from severe path loss may also greatly benefit from beamforming over a large number of antennas. Extensions to multiple-input multiple-output (MIMO) systems, or to multi-user scenarios with different performance measures (such as ergodic capacity) will thus be left as future work. In fact, after the publication of a preliminary version of this work \cite{confversion}, another paper \cite{yindijing} has studied the benefits of interleaving in hybrid single-user and multiple-user massive MIMO systems. The work \cite{yindijing} also considers a general channel model that can incorporate channel correlations. On the other hand, \cite{yindijing} ignores the feedback overhead of the interleaved scheme: It is assumed that the trained channel gains can be perfectly fed back to the transmitter, which requires an infinite number of feedback bits in practice. In contrast, we design interleaved schemes to minimize the training overhead as well as the feedback rate. In more detail, the main contributions of this paper are as follows:

\begin{itemize}
	\item We propose a novel communication scheme which interleaves training and feedback stages. In this scheme, the transmitter trains its antenna one by one while the receiver transmits the feedback information immediately after training each antenna. The feedback message may ask the transmitter to train another antenna or provide the quantized codeword to be utilized for data transmission. The latter event occurs if the already trained antennas can provide enough channel gain to avoid outage.
	
	\item We apply the interleaving scheme to a MISO system with $t$ transmit antennas, a short-term power constraint $P$, and target data rate $\rho$. We show that our scheme is able to achieve the same outage probability as a system with perfect transmitter and receiver CSI while keeping the average feedback rate and the average number of training antennas independent of $t$ and dependent only on $P$ and $\rho$.
	
	\item We design a variable-rate quantizer to minimize the feedback rate in the MISO system while keeping the same outage probability as a full-CSI system. It is achieved by allocating a higher rate to a larger coefficient in a given channel state. We also discuss the latency costs associated with interleaving, and study antenna grouping schemes as a solution.
\end{itemize}

Part of this work has been presented in a conference \cite{confversion}. Compared to \cite{confversion}, the current paper provides the proofs of technical results. It also describes a procedure to design optimal variable-rate quantizers of the feedback information. Here, we also provide numerical results that verify our analysis,  and a discussion on the latency
costs associated with interleaving and grouping. 

The rest of the paper is organized as follows: In Section \ref{secPrelim}, we describe the system model, and the full-CSI and open-loop systems. In Section \ref{secInterlelele}, we introduce the idea of interleaving and construct a simple interleaved scheme based on antenna selection. In Section \ref{secFullCSIGG}, we show how to design an interleaved scheme that can achieve the full-CSI gains with low training length and low feedback rate. In Section \ref{secVariableRateQuantizer}, we describe a variable-rate quantizer to further reduce the feedback rate. In Section \ref{seclatency}, we discuss latency costs and study interleaved schemes with antenna grouping. Finally, we present the simulation results in Section \ref{secSimulationResults} and conclusions in Section \ref{secConclusion}. Some of the technical proofs and extended discussions are provided in the appendices.

\emph{Notation:} $\mathbb{C}^{m\times n}$ is the set of all $m\times n$ complex matrices with $\mathbb{C}^m \triangleq \mathbb{C}^{m\times 1}$ and $\mathbb{C}\triangleq\mathbb{C}^1$. $\mathbb{I}_m$ is the $m\times m$ identity matrix, and $\mathbf{0}_{m\times n}$ is the $m\times n$ all-zero matrix. $\mathrm{CN}(\mathbf{K})$ is a circularly-symmetric complex Gaussian random vector with covariance matrix $\mathbf{K}$. $\mathrm{P}$ and $\mathrm{E}$ represent the probability and the expected value, respectively. $o$, $O$, and $\Theta$ are the standard Bachmann-Landau symbols. $\mathbf{A}^T$ and $\mathbf{A}^{\dagger}$ are the transpose and the conjugate transpose of matrix $\mathbf{A}$. $\circ$ stands for the entrywise product. For $x\in\mathbb{R}$, $x^+ \triangleq x$ if $x \geq 0$, and $x^+ \triangleq 0$, otherwise. For reader's convenience, we show the symbols that we will use for various schemes in the paper in Table \ref{tab:tableSymbols}.

\begin{table}
	\begin{center}
		\caption{Table of Symbols for Different Schemes.}
					\label{tab:tableSymbols}
		\begin{tabular}{|c|c|} 
		\hline
			\textbf{Symbol} & \textbf{Definition} \\
			\hline
			$\mathtt{F}$ & Full-CSI scheme\\
			$\mathtt{G}$ & Open-loop scheme\\
			$\mathtt{A}$ & Conventional antenna selection scheme\\
			$\mathtt{B}$ & Interleaved antenna selection scheme\\
			$\mathtt{B}'$ & Interleaved antenna selection scheme with antenna grouping \\
			$\mathtt{S}$ & Conventional beamforming scheme\\
			$\mathtt{D}$ & Interleaved beamforming scheme \\
			$\mathtt{C}_i,\,i=1,\ldots,t$ & Sub-blocks of interleaved beamforming scheme \\ \hline
		\end{tabular}
	\end{center}\vspace{-20pt}
\end{table}

\section{Preliminaries}
\label{secPrelim}
We consider a MISO system with $t$ transmit antennas. Denote the channel from transmit antenna $i$ to the receiver antenna by $h_i$, and let $\mathbf{h} = [h_1\cdots h_t]^T\in\mathbb{C}^{t}$ represent the entire channel state. We assume that $\mathbf{h}\simeq\mathrm{CN}(\mathbf{I}_t)$. The transmitted symbol $\mathbf{s}\in\mathbb{C}^{t}$ and the received symbol $y\in\mathbb{C}$ have the input-output relationship $y = \mathbf{s}^T\sqrt{P}\mathbf{h} + \eta$, where $P$ is the short-term power constraint of the transmitter, i.e., the total transmit power constraint over $t$ antennas, and the noise term $\eta\sim\mathrm{CN}(1)$ is independent of $\mathbf{h}$.

For a fixed $\mathbf{h}$, suppose that input symbol $\mathbf{s}$ is distributed as $\mathrm{CN}(\mathbf{K}^T)$, where $\mathbf{K}$ is a covariance matrix with $\mathrm{tr}(\mathbf{K}) \leq 1$. With perfect CSIR, the channel capacity under this strategy is $\log_2(1+\mathbf{h}^{\dagger}\mathbf{K}\mathbf{h}P)$ bits/sec/Hz. In this work, we consider a delay-constrained system where it is necessary and sufficient to sustain a certain fixed rate of data transmission at all times. Examples include video streaming for teleconferencing. In these so-called block-fading scenarios, averaging out a data codeword over infinitely many channel states is not feasible. The appropriate performance metric is the outage probability, which is the probability that the system will not be able to support a given target data rate \cite{outagemotiv1, outagemotiv2}. In our system, for a given target data transmission rate $\rho = \log_2(1+\alpha P)$, where $\alpha > 0$ can be chosen arbitrarily, an outage event occurs if $\log_2(1+\mathbf{h}^{\dagger}\mathbf{K}\mathbf{h}P)< \rho$, or equivalently if $\mathbf{h}^{\dagger}\mathbf{K}\mathbf{h} < \alpha$. We refer to the special case where $\mathbf{K} = \mathbf{x}\mathbf{x}^{\dagger}$ for some $\mathbf{x}\in\mathbb{C}^{t}$ with $\|\mathbf{x}\| \leq 1$ as ``beamforming,'' in which case the outage event is $|\langle \mathbf{x},\mathbf{h}\rangle|^2 < \alpha$. We assume that both the transmitter and the receiver agree upon a common transmission rate and power before any training or feedback communication takes place. This ensures that both terminals have perfect knowledge of $\alpha$. We also assume that there are no CSI estimation errors: Once a transmitter trains a particular antenna, the receiver can acquire the corresponding CSI error-free. The results of this paper will thus serve as upper bounds on the performance of systems that take into account possible errors in CSI estimation.

For a random $\mathbf{h}$, the transmitter can use different covariance matrices for different $\mathbf{h}$. Let $\mathtt{M}:\mathbb{C}^t\rightarrow\mathbb{C}^{t\times t}$ be an arbitrary mapping, so that given $\mathbf{h}$, the input symbol is distributed as $\mathrm{CN}([\mathtt{M}(\mathbf{h})]^T)$. The outage probability with $\mathtt{M}$ is $\mathtt{out}(\mathtt{M}) \triangleq \mathrm{P}(\mathbf{h}^{\dagger}\mathtt{M}\mathbf{h} < \alpha)$. For a beamforming-only system with mapping $\mathtt{N}:\mathbb{C}^t\rightarrow\mathbb{C}^t$, we define $\mathtt{out}(\mathtt{N}) \triangleq \mathrm{P}(|\langle \mathtt{N}(\mathbf{h}), \mathbf{h}\rangle|^2< \alpha)$.

With perfect CSIT and CSIR (a ``full-CSI'' system), the optimal mapping is beamforming along $\mathbf{h}$ \cite{HJbook}. In other words, the mapping $\mathtt{F}(\mathbf{h}) \triangleq \frac{\mathbf{h}}{\|\mathbf{h}\|}$ provides the minimum-possible outage probability $\mathtt{out}(\mathtt{F}) \!=\!\mathrm{P}(\|\mathbf{h}\|^2\! \leq \! \alpha)$.  With perfect CSIR but no CSIT (an ``open-loop'' system), it is shown in \cite{abbe1} that the optimal mapping is $\mathtt{G}(\mathbf{h})\triangleq \frac{1}{\kappa}\bigl(\begin{smallmatrix}
\mathbf{I}_{\kappa}&\mathbf{0}_{{\kappa} \times (t-{\kappa})}\\ \mathbf{0}_{(t-{\kappa}) \times {\kappa}}&\mathbf{0}_{(t-\kappa)\times (t - {\kappa})}
\end{smallmatrix}\bigr)$, where $\kappa \triangleq \arg\min_k\mathrm{P}(\sum_{i=1}^k |h_i|^2 < k\alpha)$. Hence, only $\kappa$ out of the $t$ antennas are used in general, and we have $\mathtt{out}(\mathtt{G}) = \mathrm{P}(\|\mathbf{h}_{\kappa}\|^2 < \kappa\alpha)$. \textcolor{black}{Note that $\kappa$ does not depend on the channel state $\mathbf{h}$. Therefore, the open-loop mapping is also independent of the channel state.}

The outage performance of communication systems in terms of their $\alpha$-asymptotic behaviors for a fixed $t$ has been studied in the literature. For example, a full-CSI and an open-loop system, with $t$ antennas, both provide a ``diversity gain'' of $t$ \cite{HJbook}. In other words,
given a fixed $t$, as $\alpha \rightarrow 0$, we have $\mathtt{out}(\mathtt{F}) \in \Theta(\alpha^t)$ and $\mathtt{out}(\mathtt{G}) \in \Theta(\alpha^t)$ so that the outage probabilities of a full-CSI and an open-loop system have the same $\alpha\rightarrow 0$ behavior. In contrast, in this work, we are primarily interested in the $t$-asymptotic behavior of outage probabilities for a fixed $\alpha$, i.e., the behavior of the system for a massive number of antennas. The following proposition, whose proof can be found in Appendix \ref{proofofasymptotes}, provides a rough characterization in this context.

\begin{figure}[t b]
	\centering
	\includegraphics[width=4.5in]{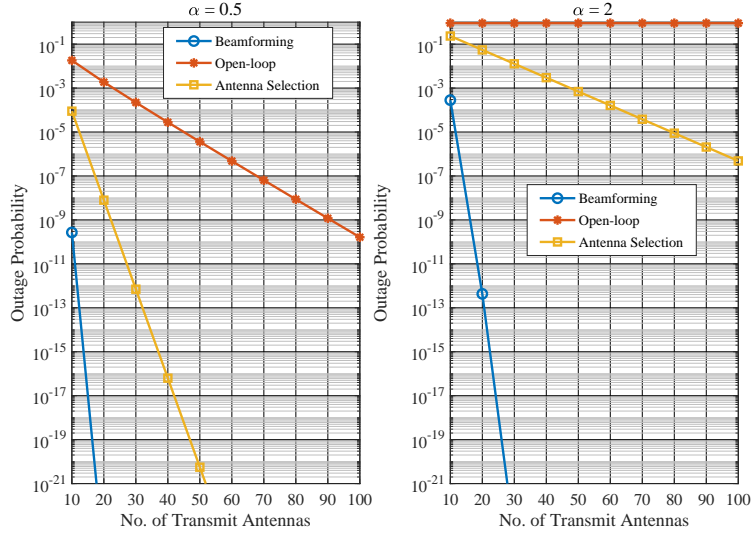}
	\caption{Outage probability as a function of the number of transmit antennas $t$ for beamforming, open-loop, and antenna selection schemes at $\alpha = 0.5$ (left) and $\alpha = 2$ (right). Note that a larger path loss exponent (due to higher frequency of transmission) or a greater transmitter-to-receiver separation translates to a higher $\alpha$ in practice. Therefore, using as many as a hundred antennas may be necessary to achieve an acceptable outage probability even in a single-user system, as evidenced by the case $\alpha =2$ and antenna selection. }
	\label{prop1}\vspace{-10pt}
\end{figure}

\begin{proposition}
\label{propasymptotes}
As $t\rightarrow\infty$, for a full-CSI system, we have $\mathtt{out}(\mathtt{F}) \in \Theta(\tfrac{\alpha^t}{t!}),\,\forall \alpha > 0$,
whereas for an open-loop system, we have $\mathtt{out}(\mathtt{G}) \in \Theta\bigl(\frac{(t\alpha)^te^{-\alpha t}}{t!}\bigr)$ if $0<\alpha < 1$, and $\mathtt{out}(\mathtt{G}) \in \Theta(1)$ if  $\alpha \geq 1$.
\end{proposition}

As is shown in Fig. \ref{prop1}, the outage probability of an open-loop system decays much \textcolor{black}{slower} than that of a full-CSI system. Proposition 1 brings both good and bad news. The good news is that for a full-CSI system, one can transmit with an arbitrarily large data rate (by choosing a sufficiently large $\alpha$) with a fixed power consumption $P$ and zero outage as $t\rightarrow\infty$. The bad news is that it is not always possible to do the same in an open-loop system: When $\alpha \geq 1$, the outage probability does not decay to $0$ with increasing $t$, and in fact, it saturates to a certain non-zero value. Also, for $0 < \alpha < 1$, even though we have $\mathtt{out}(\mathtt{G}) \rightarrow 0$ as $t\rightarrow\infty$, there is still room for improvement: As $t$ increases, the outage probability of a full-CSI system decays much faster than that of an open-loop system.

In order to obtain a vanishing outage probability as $t\rightarrow\infty$ for every $\alpha$, one should thus utilize CSIT. The full-CSI system is impractical as it requires an ``infinite'' rate of feedback from the receiver to the transmitter. A more practical approach is to settle for quantized CSIT via finite-rate receiver feedback \cite{narula1}. Another issue that is common to both a full-CSI and an open-loop system is the requirement of perfect CSIR, which may, by itself, not be feasible when $t$ is large. In the following, we thus consider the design of partial CSIT, partial CSIR schemes that interleave the training and feedback processes as shown in Fig. \ref{figinterleaved}.

%

\section{Interleaved Training and Limited Feedback}
\label{secInterlelele}
We begin with a simple example of an interleaved scheme that is based on antenna selection. We first describe its conventional non-interleaved counterpart.
\subsection{The Conventional Antenna Selection Scheme}
\label{convantenna}
A well-known partial-CSIT scheme is what we shall refer to as the ``conventional'' antenna selection scheme: Given $\mathbf{h}$, the transmitter first trains all of its antennas 
so that the receiver acquires the entire CSI. The receiver determines the antenna index $\tau \triangleq \arg\max_i |h_i|$ with the highest channel gain and sends $\lceil \log_2 t \rceil$ feedback bits to the transmitter that can uniquely represent $\tau$. The transmitter recovers $\tau$ from the feedback bits and transmits over antenna $\tau$.

This scheme can be characterized by the mapping $\mathtt{A}(\mathbf{h}) \triangleq \mathbf{e}_{\tau}$, where $\mathbf{e}_i = [\mathbf{0}_{1\times (i-1)}\,\,1\,\,\mathbf{0}_{1\times (t-i)}]^T,\,$ $i=1,\ldots,t$ are the standard basis vectors for $\mathbb{C}^t$. We have $\mathtt{out}(\mathtt{A}) = (1-e^{-\alpha})^t$, which implies $\forall\alpha>0,\,\lim_{t\rightarrow\infty} \mathtt{out}(\mathtt{A}) = 0$. Hence, for every $\alpha > 0$, we can obtain a vanishing outage probability as $t\rightarrow\infty$, as desired, which is also shown in Fig. \ref{prop1}. Moreover, for any $\alpha$ and $t$, we have $\mathtt{out}(\mathtt{A}) \leq \mathtt{out}(\mathtt{G})$, and in fact, it can be shown (e.g. by applying Stirling's approximation to the asymptotic formulae in Proposition \ref{propasymptotes}) that $\mathtt{out}(\mathtt{A}) \in o(\mathtt{out}(\mathtt{G})),\,\forall \alpha \in (0,1)$. Hence, relative to an open-loop system, antenna selection improves the $t$-asymptotic behavior of the outage probability for all $\alpha > 0$. This is shown for two values of $\alpha$ in Fig. \ref{prop1}. On the other hand, to implement this scheme, one needs to train $t$ scalar channels (one for each $h_i$) and feed back $\lceil \log_2 t \rceil$ bits for every channel. Clearly, this is not feasible in the $t\rightarrow\infty$ regime.

\subsection{A New Antenna Selection Scheme}
\label{anewantselsch}
The conventional antenna selection scheme is excessively precise in the sense that it always tries to select the antenna with the highest gain. On the other hand, without any loss of optimality in terms of the outage probability, we can in fact select \textit{any} one of the antennas that avoids outage (not necessarily \textit{the} antenna that provides the highest channel gain) whenever there is one. We use this observation to design an alternate antenna selection scheme that is based on the idea of interleaving training and limited feedback.

\begin{figure}[h]
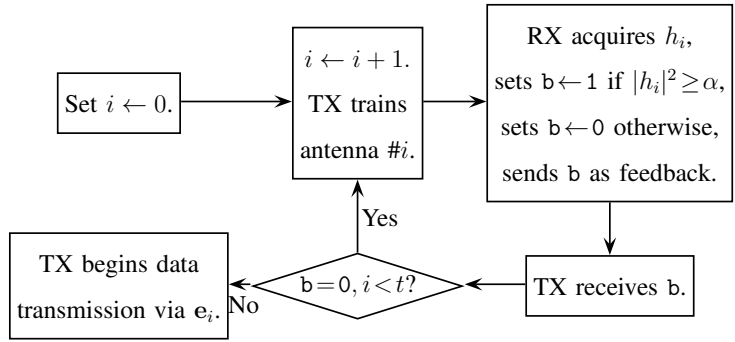

	\centering
	\scalebox{0.8}{
	\psset{shadowcolor=black!70,blur=false}
	\begin{psmatrix}[rowsep=0.5,colsep=0.36,nodealign=false]
		\psframebox[shadow=false]{\begin{tabular}{{@{}c@{}}} Set $i\leftarrow 0$. \end{tabular}} &
		\psframebox[shadow=false]{\begin{tabular}{{@{}c@{}}} $i\leftarrow i + 1.$ \\ TX trains \\ antenna \#$i$. \end{tabular}} &
		\psframebox[shadow=false]{\begin{tabular}{{@{}c@{}}} RX acquires $h_i$,  \\ sets $\mathtt{b}\!\leftarrow\! \mathtt{1}$ if $|h_i|^2\! \geq \! \alpha$, \\ sets $\mathtt{b}\!\leftarrow\! \mathtt{0}$ otherwise, \\ sends $\mathtt{b}$ as feedback. \end{tabular}}  \\
		\psframebox[shadow=false]{\begin{tabular}{{@{}c@{}}}  TX begins data \\transmission via $\mathbf{e}_i$. \end{tabular}} & \psdiabox[shadow=false]{\hspace{-4.5pt}$\mathtt{b}\!=\!\mathtt{0}, i\!<\! t$?\hspace{-5pt}}
		&\psframebox[shadow=false]{\begin{tabular}{{@{}c@{}}} TX receives $\mathtt{b}$. \end{tabular}}  &
		\psset{arrows=->,fillstyle=none, arrowsize=6pt}\ncline{1,1}{1,2}\ncline{1,2}{1,3}\ncline{1,3}{2,3}\ncline{2,2}{1,2}>{\!\!\!\!\!\begin{tabular}{c} \\ Yes \end{tabular}}\ncline{2,2}{2,1}_{\hspace{5pt}No}
		\psset{arrows=->,fillstyle=none,arrowsize=6pt}
		\ncline{2,3}{2,2}
	\end{psmatrix}
	}
\caption{The new antenna selection scheme. Note that the variable $i$, i.e., the antenna number, can be thought to be ``naturally available'' to both the transmitter and the receiver: At both terminals, it can be initialized and updated throughout the multiple training and feedback stages without any extra overhead. Also note that the receiver is always aware of what the next action (training a new antenna or beginning the data transmission) of the transmitter is going to be so that there is no inconsistency. This is because it is the receiver itself that provides the feedback message, which uniquely determines the  transmitter action. }
\label{figmodantennaselect}\vspace{-10pt}
\end{figure}

Our new antenna selection scheme operates as shown in Fig. \ref{figmodantennaselect}: The transmitter first trains the channel $h_1$ corresponding to the first antenna and waits for receiver feedback. The receiver, having acquired the knowledge of $h_1$, sends the one-bit feedback message ``$\mathtt{1}$'' if $|h_1|^2 \geq \alpha$, i.e. if selecting the first antenna avoids outage. Otherwise, it feeds back a ``$\mathtt{0}$,'' which indicates that selecting the first antenna will result in an outage. Now, if the transmitter receives a ``$\mathtt{1}$,'' the training and feedback process can end; the transmitter starts data transmission over the first antenna only (without the need of training the remaining antennas) and outage is avoided. Otherwise, if the transmitter receives a ``$\mathtt{0}$,'' it proceeds to training the channel state $h_2$ corresponding to its second antenna. The process continues in the same manner until an antenna (selection vector) that avoids outage is found. If all the antennas result in an outage, then the transmitter can simply transmit over an arbitrary antenna.

Clearly, the new scheme achieves the same outage probability $(1-e^{-\alpha})^t$ as the conventional scheme discussed in Section \ref{convantenna}. Now, given $1\leq i \leq t-1$, the transmitter trains only the first $i$ antennas with probability $e^{-\alpha}(1-e^{-\alpha})^{i-1}$, and it trains all the $t$ antennas with probability $(1-e^{-\alpha})^{t-1}$. The training length, which we define as the average number of antennas that are trained per channel state, is thus $$\textstyle \sum_{i=1}^t ie^{-\alpha}(1-e^{-\alpha})^{i-1} + t(1-e^{-\alpha})^t = e^{\alpha}(1-(1-e^{-\alpha})^t).$$
A similar calculation reveals that the feedback rate of the scheme, which we define as the average number of bits that are fed back per channel state, is actually (numerically) equal to its training length. Hence, the training and the feedback rates of the new scheme are both given by the formula $e^{\alpha}(1-(1-e^{-\alpha})^t)$. Note that for any $t$, the two rates are both upper bounded by $e^{\alpha}$, which is independent of $t$.

The significance of the new scheme is that it provides a vanishing outage probability as $t\rightarrow\infty$ with $t$-independent training length and feedback rate. One can thus obtain the benefits of having infinitely many antennas with finite training and feedback overheads. For example, setting $\alpha = 1$, we can observe that if the transmitter has infinitely many antennas, then for any given power constraint $P$, we can transmit with rate $\log(1+P)$ bits/sec/Hz outage-free via training only $e < 3$ antennas and feeding back $3$ bits on average. Comparison with an open-loop system (a system with perfect CSIR but no CSIT) leads to the following conclusion: It is much better to have a little bit of CSIT and a little bit of CSIR rather than to have perfect CSIR but no CSIT.

We note that our interleaved antenna selection scheme can also be applied to the orthogonal frequency division multiplexing (OFDM) systems. The main challenge is that the best selection of antennas is likely to change with frequency. As is shown in \cite{xgaoOFDMantSelection}, the antenna selection problem can be formulated as finding the antenna with the best channel averaged over all sub-carriers. As a result, we may use the average channel gain over all sub-carriers to determine whether a specific antenna is outage-avoiding or not.

\newcommand{\variationsofintleaving}{Several variations on our interleaved antenna selection scheme can be considered. For example, in order to avoid the possible implementation complexities and delays of training the antennas one by one, the transmitter may train all $t$ antennas at once as in conventional antenna selection. On the other hand, the receiver may now use variable-length feedback instead of the $\lceil \log_2 t \rceil$ bits of fixed length feedback in conventional antenna selection. In detail, suppose that selecting any of the first $\upsilon$ antennas results in an outage, but selecting Antenna $\upsilon+1$ avoids outage, where $\upsilon\in\{0,\ldots,t\}$. We let $\upsilon = t$ if selecting any of the $t$ antennas results in outage. The receiver then feeds back the binary codeword $1\cdots 1 0$, where there are $\upsilon$ ones. The transmitter can recover the outage-avoiding antenna from the feedback information if such an antenna exists. This scheme, which utilizes fixed-length training and variable-length feedback, lies in between the two extremes of conventional antenna selection (that uses fixed-length training and feedback), and interleaved antenna selection (that uses variable-length training and feedback). It is a special case of the variable-length beamforming schemes in \cite{koyuncu1} for full-CSIR systems. It achieves the same outage probability as conventional antenna selection with training length $t$, and feedback rate $e^{\alpha}(1-(1-e^{-\alpha})^t)$. Note that the feedback rate of the scheme equals that of interleaved antenna selection and thus remains bounded as $t\rightarrow\infty$. As discussed in \cite{koyuncu1}, the feedback rate  may possibly be reduced further with a better codeword assignment; e.g., by using Huffman's algorithm. Nevertheless, the training length of the scheme grows without bound as $t\rightarrow\infty$.  Later in Section \ref{seclatency}, we shall consider other variations that rely on training a subset of antennas at a time instead of training all antennas at once or training them one by one.}\variationsofintleaving

\subsection{General Description of an Interleaved Scheme}
\label{secgendescintl}
So far, we have discussed many seemingly-different scenarios including non-interleaved or interleaved schemes, the full-CSI and the open-loop systems, and so on. All of these scenarios can in fact be viewed as manifestations of a single unifying framework of a \textit{generalized beamforming scheme}, which describes the rules of how the tasks of training and feedback are to be performed. The advantage of this viewpoint is that it will allow us to more meaningfully compare different scenarios with respect to their outage probabilities, training lengths, and feedback rates. We call this generalized beamforming scheme, as defined below, Scheme $\mathtt{S}$.

One task of Scheme $\mathtt{S}$ is to specify the quantized covariance matrix $\mathtt{S}(\mathbf{h})$ to be utilized given channel state $\mathbf{h}$. By the definitions in Section \ref{secPrelim}, the outage probability with $\mathtt{S}$ is thus given by  $\mathtt{out}(\mathtt{S})$. Scheme $\mathtt{S}$ also describes which antennas are to be trained in which order, the corresponding feedback messages of the receiver, and how these messages are decoded at the transmitter. Obviously, different choices result in different schemes and different performances. An example of these ``inner workings'' of Scheme $\mathtt{S}$ can be found in Section \ref{anewantselsch} for the special case of our new antenna selection scheme. As such, while we use Scheme $\mathtt{S}$ to represent the general structure of our beamforming scheme, when the details of training, feedback, transmission and decoding are defined, i.e., a specific scheme is defined in details as done in Section \ref{anewantselsch}, we will use a specific name for the specific scheme.  The two important figures of merit of Scheme $\mathtt{S}$ is its training length $\mathtt{tl}(\mathtt{S})$ and its feedback rate $\mathtt{fr}(\mathtt{S})$, which can be defined in the same manner as we have done in Section \ref{anewantselsch}.


We can now view a full-CSI system, called Scheme $\mathtt{F}$, as an example of Scheme $\mathtt{S}$. Operationally, a full-CSI system trains all its antennas and performs the optimal beamforming along the direction $\frac{\mathbf{h}}{\|\mathbf{h}\|}$. As a result, we will have $\mathtt{out}(\mathtt{F}) = \mathrm{P}(\|\mathbf{h}\|^2 < \alpha)$ and $\mathtt{tl}(\mathtt{F}) = t$. Since representing an arbitrary beamforming vector requires an infinite rate of feedback, we have $\mathtt{fr}(\mathtt{F}) = \infty$. Similarly, the open-loop scheme $\mathtt{G}$ trains the first $\kappa$ antennas. Since there is no feedback, $\mathtt{fr}(\mathtt{G}) = 0$ and the transmitter sends independent Gaussian symbols with equal energy over the first $\kappa$ antennas. Therefore, we have  $\mathtt{out}(\mathtt{G}) = \mathrm{P}(\|\mathbf{h}_{\kappa}\|^2 < \kappa \alpha)$ and
$\mathtt{tl}(\mathtt{G}) = \kappa$. Also, as shown in Section \ref{convantenna}, the conventional antenna selection system, called Scheme $\mathtt{A}$, will have $\mathtt{out}(\mathtt{A}) = (1-e^{-\alpha})^t$, $\mathtt{tl}(\mathtt{A}) = t$, and $\mathtt{fr}(\mathtt{A}) = \lceil \log_2 t \rceil$.

Clearly, Scheme $\mathtt{S}$ provides a framework to extend the previous definitions in a consistent manner and offers a set of quantities to compare the performance of different schemes. For example, we can summarize the performance metrics of our new antenna selection scheme in Section \ref{anewantselsch}, called Scheme $\mathtt{B}$, in the following theorem:

\begin{theorem}\label{Thm1}
Scheme $\mathtt{B}$, defined in Section \ref{anewantselsch}, provides $\mathtt{out}(\mathtt{B}) = \mathtt{out}(\mathtt{A}) = (1-e^{-\alpha})^t$ and $\mathtt{tl}(\mathtt{B}) = \mathtt{fr}(\mathtt{B})  = e^{\alpha}(1-(1-e^{-\alpha})^t) < e^{\alpha}$.
\end{theorem}

These results lead to the following question: What is the best-possible outage probability for given constraints on training length and feedback rate? Unfortunately, this problem appears to be difficult in general, and we thus leave a detailed treatment as future work. In a related direction, Theorem \ref{Thm1} shows the existence of a ``good'' scheme that can achieve a vanishing outage probability as $t\rightarrow\infty$ with $t$-independent feedback and training lengths. One fundamental question that immediately comes to mind is then to determine whether one can achieve the ultimate limit $\mathtt{out}(\mathtt{F})$ with again $t$-independent training length and feedback rate. The answer is yes, and the construction of such a scheme will be provided next. Meanwhile, we note that even though antenna selection provides a reasonable performance, we still have $\mathtt{out}(\mathtt{F}) \in o(\mathtt{out}(\mathtt{A}))$ as $t\rightarrow\infty$. In other words, the outage probability with a full-CSI system decays much faster than the one with antenna selection.
While we have shown this fact analytically, Fig. \ref{prop1} demonstrates it numerically as well. This also provides a ``practical motivation'' for construction of schemes that achieve the full-CSI gains.

\section{Achieving the Full-CSI Gains by Interleaving}
\label{secFullCSIGG}
Our construction here relies on our earlier work \cite{koyuncu1}, which introduced the idea of variable-length feedback for a MISO system with perfect CSIR. We thus first recall some of the relevant technical tools and results.

\subsection{Variable-Length Limited Feedback with Perfect CSIR}
\label{originalvlq}
We begin by defining a simple deadzone scalar quantizer. For any given integer $\ell\geq 0$ and $x\in[-1,+1]$, let $q(x;\ell) \triangleq \mathrm{sign}(x)\frac{1}{2^{\ell+1}}\lfloor |x|2^{\ell+1} \rfloor$. We can easily calculate $q(x; \ell)$ by taking the most significant $\ell + 2$ bits $(b_0.b_1b_2\cdots b_{\ell+1})_2$ of the binary representation $(b_0.b_1b_2\cdots)_2$ of $|x|$, while preserving the sign of $x$. For example, we have $q(\pm(0.101)_2; 1) = \pm(0.10)_2$.

We extend the definition of the deadzone quantizer $q$ to an arbitrary beamforming vector $\mathbf{x} = [x_1\cdots x_t]^T \! \in\! \mathbb{C}^{t}$ with $\|\mathbf{x}\|\leq 1$ by setting $q(\mathbf{x};\ell) \triangleq [\,q(\Re x_1;\ell) + j q(\Im x_1; \ell) $ $ \,\cdots\, q(\Re x_t;\ell) + j q(\Im x_t; \ell)\,]^T\in\mathbb{C}^{t}$. We refer to the parameter $\ell$ as the ``resolution'' of $q$. Note that by construction, $\|q(\mathbf{x};\ell)\| \leq 1$, and therefore, $q(\mathbf{x};\ell)$ is itself a feasible beamforming vector. Moreover, for a fixed $\ell$ and $t$, each quantized vector $q(\mathbf{x};\ell)$ can be uniquely represented by $2t(\ell + 3)$ bits (For each of the $2t$ complex dimensions of $\mathbf{x}$, we spend one bit for the sign, and $\ell + 2$ bits for the most significant $\ell + 2$ binary digits.).

Now, for an arbitrary channel state $\mathbf{h}$ with $\|\mathbf{h}\|^2 > \alpha$, let
$L(\mathbf{h})\!\triangleq\! \max\{\lceil \log_2(4t) \rceil, \lceil \log_2\frac{4t\alpha}{\|\mathbf{h}\|^2-\alpha}\rceil\}$,
and $\vech \triangleq \mathtt{F}(\mathbf{h}) = \frac{\mathbf{h}}{\|\mathbf{h}\|}$. We have the following proposition.
\begin{proposition}[{{\!\!\cite[Proposition 4]{koyuncu1}}}\hspace{0.75pt}]
\label{wwprop}
Let $\mathbf{h}\in\mathbb{C}^{t}$ with $\|\mathbf{h}\|^2 > \alpha$ for some $t \geq 1$.
Then,
\begin{align}
|\langle q(\vech;L(\mathbf{h})), \mathbf{h} \rangle|^2 > \alpha.
\end{align}
\end{proposition}
This result has the following interpretation. Suppose $\|\mathbf{h}\|^2 > \alpha$, and thus outage is avoidable with the beamforming vector $\vech$. By construction, the sequence of quantized beamforming vectors $q( \vech;\ell),\,\ell\geq 0$ (which are feasible since $\|q( \vech;\ell)\| \leq \| \vech\| = 1$) provides an increasingly finer approximation of $\vech$ as the resolution $\ell$ grows to infinity. The proposition shows that for every given $\mathbf{h}$ with $\|\mathbf{h}\|^2 > \alpha$, there is in fact a ``sufficient resolution'' $L(\mathbf{h})$ (that depends only on $\|\mathbf{h}\|$) such that the quantized beamforming vector $q( \vech;\ell)$ can avoid outage.

As discussed in \cite{koyuncu1}, Proposition \ref{wwprop} leads to the following limited feedback scheme under the assumption of perfect CSIR: If $\|\mathbf{h}\|^2 > \alpha$, the receiver calculates the required resolution $L(\mathbf{h})$ to avoid outage, and sends $2t(L(\mathbf{h}) + 3)$ feedback bits that represent the corresponding outage-avoiding beamforming vector $q(\vech;L(\mathbf{h}))$. The transmitter, which we assume can perfectly know the length of the feedback codeword that it has received, first recovers $L(\mathbf{h})$, and then the beamforming vector $q(\vech;L(\mathbf{h}))$. Otherwise, if $\|\mathbf{h}\|^2 \leq \alpha$, outage is unavoidable except for channel states $\|\mathbf{h}\|^2 = \alpha$ with zero probability. In this case, the receiver sends the one-bit feedback message ``$\mathtt{0}$'' so that the transmitter can transmit with an arbitrary but fixed beamforming vector, say $\mathbf{e}_1$. We refer to this scheme as Scheme $\mathtt{C}_t$, where the subscript indicates the number of transmit antennas. We have $\mathtt{C}_t(\mathbf{h}) \!=\! q(\vech;L(\mathbf{h}))$. By construction, Scheme $\mathtt{C}_t$ achieves the full-CSI outage probability with the feedback rate
\begin{align}
\label{fratenomrlavarrate}
\mathtt{fr}(\mathtt{C}_t) = \mathrm{P}(\|\mathbf{h}\|^2 \leq \alpha) + \textstyle\sum_{\ell = \lceil \log_2 (4t) \rceil }^{\infty} 2t(\ell + 3)p_{\ell},
\end{align}
where $p_{\ell} \triangleq \mathrm{P}(L(\mathbf{h}) = \ell,\|\mathbf{h}\|^2 > \alpha)$. As $\ell\rightarrow\infty$,  $p_{\ell}$ can be shown to decay fast enough so that the resulting feedback rate is finite; we refer the interested reader to \cite{koyuncu1} for the details and formal calculations. Intuitively, instead of trying to pick the best beamforming vector that maximizes the signal-to-noise ratio in some given codebook, one spends just enough bits to describe a beamforming vector that avoids outage. This allows us to achieve the full-CSI performance with a finite feedback rate under the assumption of perfect CSIR.
\subsection{Achieving $\mathtt{out}(\mathtt{F})$ by Interleaving}
We now return to our main goal of designing a scheme that can achieve the full-CSI outage probability with finite training length and feedback rate. Scheme $\mathtt{C}_t$ as described above is not immediately applicable for our purposes as (i) it requires perfect CSIR and thus induces a training length of $t$, and (ii) according to (\ref{fratenomrlavarrate}), its feedback rate grows at least as $\Theta\hspace{-1pt}(t)$ (We have $\mathtt{fr}(\mathtt{C}_t) \geq 6t \sum_{\ell = \lceil \log_2 (4t) \rceil }^{\infty} p_{\ell} = 6t\mathrm{P}(\|\mathbf{h}\|^2 > \alpha)  \in \Theta(t)$.).

\begin{figure}[h]
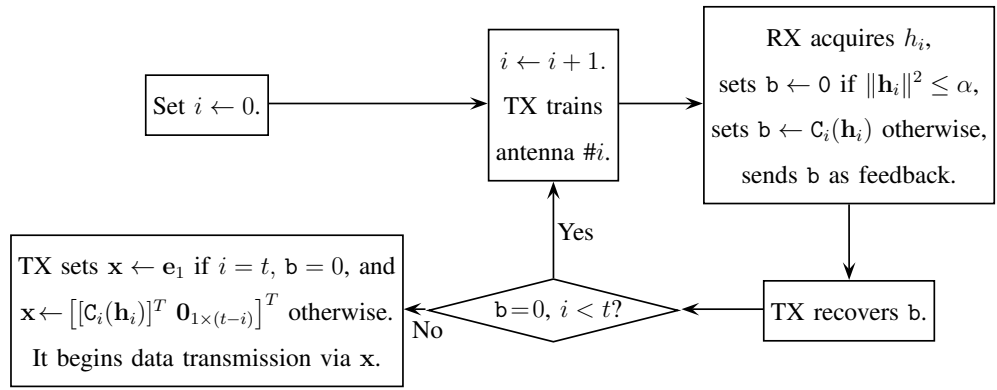

	\centering
	\scalebox{0.8}{
	\psset{shadowcolor=black!70,blur=false}
	\begin{psmatrix}[rowsep=0.5,colsep=0.36,nodealign=false]
		\psframebox[shadow=false]{\begin{tabular}{{@{}c@{}}} Set $i\leftarrow 0$. \end{tabular}} &
		\psframebox[shadow=false]{\begin{tabular}{{@{}c@{}}} $i\leftarrow i + 1.$ \\ TX trains \\ antenna \#$i$. \end{tabular}} &
		\psframebox[shadow=false]{\begin{tabular}{{@{}c@{}}} RX acquires $h_i$, \\ sets  $\mathtt{b}\leftarrow \mathtt{0}$ if $\|\mathbf{h}_i\|^2 \leq \alpha$, \\ sets $\mathtt{b} \leftarrow \mathtt{C}_i(\mathbf{h}_i)$ otherwise, \\ sends $\mathtt{b}$ as feedback. \end{tabular}}  \\
		\psframebox[shadow=false]{\begin{tabular}{{@{}c@{}}} TX sets $\mathbf{x}\leftarrow\mathbf{e}_1$  if $i=t,\,\mathtt{b}=0$, and \\ $\mathbf{x}\!\leftarrow\! \bigl[[\mathtt{C}_i(\mathbf{h}_i)]^T\,\, \mathbf{0}_{1\times (t-i)}\bigr]^T$ otherwise. \\ It begins data transmission via $\mathbf{x}$. \end{tabular}} & \psdiabox[shadow=false]{\hspace{-4.5pt} {$\mathtt{b}\!=\!0$, $i < t$}?\hspace{-5pt}}
		&\psframebox[shadow=false]{\begin{tabular}{{@{}c@{}}} TX recovers $\mathtt{b}$. \end{tabular}}  &
		\psset{arrows=->,fillstyle=none, arrowsize=6pt}\ncline{1,1}{1,2}\ncline{1,2}{1,3}\ncline{1,3}{2,3}\ncline{2,2}{1,2}>{\!\!\!\!\!\begin{tabular}{c} \\ Yes \end{tabular}}\ncline{2,2}{2,1}_{\hspace{45pt}No}
		\psset{arrows=->,fillstyle=none,arrowsize=6pt}
		\ncline{2,3}{2,2}
	\end{psmatrix}}
	\caption{Operation of scheme $\mathtt{D}$. Due to the equivalence between
		$\mathtt{C}_i(\mathbf{h}_i)= q(\vechi;L(\mathbf{h}_i))$ and its binary description (see Section IV.A), we use the same notation ``$\mathtt{C}_i(\mathbf{h}_i)$'' for the codeword of $2i(L(\mathbf{h}_i)+3)$ bits that represent $\mathtt{C}_i(\mathbf{h}_i)$. }
	\label{figbeamfoorming}\vspace{-10pt}
\end{figure}

We can however incorporate the sequence of Schemes $\mathtt{C}_i,\,i=1,\ldots,t$ as sub-blocks  of an interleaved training and limited feedback Scheme $\mathtt{D}$ as shown in Fig. \ref{figbeamfoorming}. In the figure, we use the notation $\mathbf{h}_i \triangleq [h_1\cdots h_i]^T,\,i=1,\ldots,t$ to represent the first $i$ components of the channel state $\mathbf{h}$. Given $\mathbf{h}$ and a value of the variable $i\in\{1,\ldots,t\}$ in the figure, suppose that the transmitter has ``just'' trained its $i$th antenna, so that the receiver has acquired the knowledge of $h_i$. At this stage, the receiver knows the channel values $h_1,\ldots,h_i$ corresponding to the first $i$ antennas of the transmitter, or equivalently, it knows $\mathbf{h}_i$. We consider the following two cases for the receiver's feedback and the corresponding transmitter action.

If $\|\mathbf{h}_i\|^2 \leq \alpha$, as far as the channels that have been made available to the receiver are concerned, outage is unavoidable with probability $1$. The receiver thus requests the transmitter to train the next antenna by sending the feedback bit ``$\mathtt{0}$,'' and the transmitter complies. The case $i=t$ is an exception: Outage is unavoidable with any beamforming vector with probability $1$ (we have $\|\mathbf{h}_t\|^2 = \|\mathbf{h}\|^2 \leq \alpha$), and thus the transmitter transmits via the (arbitrarily chosen) vector $\mathbf{e}_1$.

On the other hand, if $\|\mathbf{h}_i\|^2 > \alpha$, the receiver feeds back the $i$-dimensional vector $\mathtt{C}_i(\mathbf{h}_i) = q(\vechi;L(\mathbf{h}_i))$ using $2i(L(\mathbf{h}_i) + 3)$ feedback bits. By Proposition \ref{wwprop}, we have $|\langle \mathtt{C}_i(\mathbf{h}_i) , \mathbf{h}_i \rangle|^2 > \alpha$. This implies that the actual $t$-dimensional beamforming vector utilized at the transmitter, which is simply constructed by appending $t-i$ zeroes to $\mathtt{C}_i(\mathbf{h}_i) $, will also avoid outage.

By construction, Scheme $\mathtt{D}$ avoids outage for any channel state $\mathbf{h}$ with $\|\mathbf{h}\|^2 > \alpha$. Hence, it achieves the full-CSI outage probability $\mathtt{out}(\mathtt{F})$. Calculations for the training length and feedback rate of Scheme $\mathtt{D}$ are slightly more involved. We present the final results by the following theorem, whose proof can be found in Appendix \ref{proofofschbftheo}.
\begin{theorem}\label{Thm2}
We have $\mathtt{out}(\mathtt{D}) = \mathtt{out}(\mathtt{F})$ with $\mathtt{tl}(\mathtt{D}) \leq 1+\alpha$ and $\mathtt{fr}(\mathtt{D}) \leq 92(1+\alpha^3)$.
\end{theorem}
We shall emphasize that Theorem \ref{Thm2} should be interpreted as ``just'' an achievability result. Its main message is that the full-CSI performance can be achieved with $t$-independent training length and feedback rate. Hence, the $\alpha$-dependent bounds in the statement of Theorem \ref{Thm2} are not necessarily the best-possible as far as a general scheme that can achieve $\mathtt{out}(\mathtt{F})$ is concerned. As can be observed from the proof of the theorem, we have not tried to optimize the bounds. 

Let us now also compare the results of Theorem \ref{Thm2} with what we have achieved by Theorem \ref{Thm1} using the antenna selection Scheme $\mathtt{B}$. For Scheme $\mathtt{B}$, we have $\mathtt{tl}(\mathtt{B}),\mathtt{fr}(\mathtt{B}) \in  \Theta(e^{\alpha})$ as $\alpha \rightarrow \infty$, while for Scheme $\mathtt{D}$, we have $\mathtt{tl}(\mathtt{D}) \in O(\alpha)$ and $\mathtt{fr}(\mathtt{D}) \in O(\alpha^3)$.  Hence, there are certain values of $t$ and $\alpha$ where Scheme $\mathtt{D}$ improves upon Scheme $\mathtt{B}$ in every aspect. It should be clear why Scheme $\mathtt{D}$ provides a better outage performance. Regarding the training lengths, note that Scheme $\mathtt{B}$ terminates only if the \textit{most-recently} trained antenna avoids outage. On the other hand,  Scheme $\mathtt{D}$ terminates whenever the joint contribution of \textit{all} trained antennas avoids outage. Therefore, for every channel state, Scheme $\mathtt{D}$ always terminates before Scheme $\mathtt{B}$ does, and thus, in fact, $\mathtt{tl}(\mathtt{D}) \leq \mathtt{tl}(\mathtt{B})$. The efficiency of Scheme $\mathtt{D}$ in terms of training also positively affects its feedback rate: The fewer the amount of antennas that one needs to train, the fewer the feedback messages spent requesting these antennas to be trained. \textcolor{black}{In both cases, same outage probability results in the same diversity.}

An interesting special case of Theorem \ref{Thm2} is to assume $P$ is large (but still fixed), and choose $\alpha = P^{m-1}$ for some $m>1$. Then, if the transmitter has infinitely many antennas (for a simpler discussion, we put the physical impossibility of such an assumption aside), Theorem \ref{Thm2} tells us that we can transmit with rate $\log(1+P^m) \sim m\log P$ (as $P\rightarrow\infty$) outage-free, and thus achieve a multiplexing gain of $m$. In other words, one can achieve ``the MIMO effect'' from a MISO system with a very large number of antennas. The price to pay however is a training length of $O(P^m)$ and a feedback rate of $O(P^{3m})$, which are both much larger than the data transmission rate $m\log P$. Ideally, we would like the feedback and training lengths in Theorem \ref{Thm2} (or in another scheme with a $t\rightarrow\infty$ vanishing outage probability) to be $o(\log \alpha)$ as $\alpha \rightarrow \infty$. Whether this is possible or not will remain as an interesting open problem and shows the need for proving converse results for general interleaved schemes.

On the other hand, regarding the data rate $\log(1+\alpha P)$, when $P$ is small (a typical case of a low-power system), even slight increase in $\alpha$ significantly improves the data transmission rate. For example, for $P=1$, increasing $\alpha$ from $1$ to $3$ doubles the data rate. For such scenarios with small $P$, tighter bounds on the training lengths, feedback rates and/or custom-made numerically-designed interleaved schemes are a necessity. In this context, tighter bounds are desirable as they will provide a more accurate estimate on the required training and/or feedback rates to achieve a certain outage probability. On the other hand, numerical designs are desirable as they may outperform the analytically-constructed schemes. Finding an efficient algorithm for the numerical design of interleaved schemes would prove to be a challenging network vector quantization problem \cite{fleming1}, where one has to design several interdependent vector quantizers managing the multiple feedback phases of the interleaved scheme. In particular, given $t$ transmitter antennas, one has to design $t$ vector quantizers, $Q_1,\ldots,Q_t$, where the domain of $Q_i$ depends the range of $Q_{i-1}$. An alternating optimization approach may then be taken where, for the infinite sequence $i=1,\ldots,t,1,\ldots,t,\ldots$, one optimizes $Q_i$ while fixing $Q_j,\,j\neq i$.

\section{Quantization Rate Allocation}
\label{secVariableRateQuantizer}
We now discuss how to further reduce the feedback rate of our proposed schemes using an optimized rate allocation  strategy. Recall that in the construction in Section \ref{originalvlq}, one spends a fixed $2(L(\mathbf{h}) + 3)$ bits per antenna to encode each component of the beamforming vector. Different components of a beamforming vector have different weights in the array gain which is given as $|\langle q(\vech;L(\mathbf{h})), \mathbf{h} \rangle|^2$. A component with higher weight should be quantized more accurately, i.e., assigned a higher rate, to provide a better overall performance \cite{quantizationGray}.

For a given beamforming vector $\mathbf{x}$, we assign the optimal quantization rate to each component. To accommodate a variable-rate for different components, we need to adjust the resolution $\ell$ of the deadzone quantizer. Instead of using the fixed resolution $\ell$ for all components, resulting in a fixed-rate system, we use the resolution $\ell_{ij}$ ($i=1,\cdots, t,\ j=1, 2$) for the real (if $j = 1$) or imaginary (if $j = 2$) part of $x_i$. This will result in a variable-rate deadzone quantizer $q_v$ to be defined for an arbitrary beamforming vector $\mathbf{x} = [x_1 \cdots x_t]^T \in\mathbb{C}^{t}$ with $\lVert \mathbf{x} \rVert \leqslant 1$ as $q_v(\mathbf{\mathbf{x}; {\boldsymbol\ell}}) \triangleq [\,q(\Re x_1;\ell_{11}) + j q(\Im x_1; \ell_{12}) $ $ \,\cdots\, q(\Re x_t;\ell_{t1}) + j q(\Im x_t; \ell_{t2})\,]^T\in\mathbb{C}^{t}$, where $q$ is the deadzone scalar quantizer and $\boldsymbol\ell$ is a $t \times 2$ matrix representing the resolution of $q$ for real and imaginary parts of different components in $\mathbf{x}$. Note that by the definition of the deadzone quantizer $q$, $\lvert q(x; \ell) \rvert \leq \lvert x \rvert $ for any $x \in [-1, 1]$ and any positive $\ell$. Therefore, $\lVert q_v(\mathbf{x}; \boldsymbol{\ell}) \rVert \leq 1$, which means $q_v(\mathbf{x}, \mathbf{\ell})$ is also a feasible beamforming vector.

\begin{algorithm}
\caption{Rate-Allocation Algorithm}
\begin{algorithmic}[1]
\State Set $\boldsymbol\ell$ to be the $k \times 2$ all-zero matrix, $\vech_k = \frac{\mathbf{h}_k}{\lVert \mathbf{h}_k \rVert}$, and $count = 0$.
\State Set $L(\mathbf{h}_k) = \max\{\lceil \log_2(4k) \rceil, \lceil \log_2\frac{4k\alpha}{\|\mathbf{h}_k\|^2-\alpha}\rceil\}$.
\While{$count < 2k(L(\mathbf{h}_k) + 3)$ \textbf{and} $|\langle q_v(\vech_k;\boldsymbol\ell), \mathbf{h}_k \rangle|^2 < \alpha$}
\State $\mathbf{e} = q_v(\vech_k;\boldsymbol\ell + \mathbf{\Delta}) - q_v(\vech_k;\boldsymbol\ell)$, where $\mathbf{\Delta} \triangleq \Delta  \mathbf{J}_{k\times 2}$, and $\mathbf{J}_{k\times 2}$ is the $k\times 2$ all-one matrix.
\State $\mathbf{d}_1 = \Re \vech_k \circ \Re \mathbf{e}$, $\mathbf{d}_2 = \Im \vech_k \circ \Im \mathbf{e}$.
\State Find the indices $i$ and $j$ corresponding to the maximum values of $\mathbf{d}_1$ and $\mathbf{d}_2$, respectively.
\State \textbf{If} $\mathbf{d}_1[i] > \mathbf{d}_2[j]$, \textbf{then} $\boldsymbol\ell[i, 1] = \boldsymbol\ell[i, 1] + \Delta$, \textbf{else}  $\boldsymbol\ell[j, 2] = \boldsymbol\ell[j, 2] + \Delta$.
\State $count = count + \Delta $.
\EndWhile
\State \Return $q_v(\vech_k; \boldsymbol\ell)$.
\end{algorithmic}
\end{algorithm}

To formulate it as a classic rate-allocation problem in a rate-distortion set-up, we define
	$R_a \triangleq \sum_{i=1}^{k}\sum_{j=1}^{2}\ell_{ij}$, and $D_a \triangleq |\langle q_v(\vech_k;\boldsymbol\ell), \mathbf{h}_k \rangle|^2 \geqslant \alpha$. The optimal rate-allocation will be achieved by assigning the appropriate quantization rate $\boldsymbol\ell$ to each component of $\vech_k$ to minimize $R_a$ while satisfying the constraint on $D_a$. This rate-allocation problem is the dual of the bit-allocation problem in data compression, which is well studied \cite{westerinkBitAllocation, shohamBitAllocation, riskinBFOS}. Typically, the bit-allocation problem is to minimize the overall distortion under some constraint on the total bit rate while the proposed rate-allocation problem is to minimize the total bit rate under some constraint on the overall distortion.
As a result, the generalized Breiman, Friedman, Olshen, and Stone (BFOS) algorithm \cite{riskinBFOS} can be utilized to solve our rate-allocation problem. We design Algorithm 1, based on the generalized BFOS algorithm in \cite{riskinBFOS}, to find the optimal rate-allocation to quantize a beamforming vector. The main idea behind the algorithm is as follows. At each step of the algorithm, we assign additional $\Delta$ bits to the beamforming vector component that results in the maximum distortion reduction among all possible vector components. This will result in an increase of $\Delta$ bits to the total quantization rate and a reduction in the total distortion, i.e., an increase in the array gain. After updating the rate and distortion of the chosen component, we continue the iterations until the overall distortion satisfies the constraint $D_a \geq \alpha$ or the total quantization rate is greater than that of the fixed-rate deadzone quantizer.

\section{Latency Considerations and Antenna Grouping}
\label{seclatency}
Our formulations so far ignore the extra latency incurred by dividing the training and feedback stages to multiple stages, as in the proposed interleaved schemes. In this section, we study the latency/performance tradeoffs of interleaving by assuming that every stage of training and interleaving consumes an extra $\epsilon$-fraction of the time that would otherwise be spent on data transmission. This $\epsilon$-cost may, for example, stem from the propagation delays between the transmitter and the receiver during the training and feedback phase.

In such a scenario, training the antennas one by one, as in the previous sections, may be too costly, and thus suboptimal. For this reason, we consider an interleaved antenna selection with antenna grouping that trains antennas $K$ by $K$, where $K \geq 1$. For simplicity, we assume $T$ is a multiple of $K$. The transmitter trains the first $K$ antennas and the receiver acquires the CSI for the first $K$ antennas $h_1,\ldots,h_K$. The receiver sends $\lceil \log_2(1+K)\rceil$ bits of feedback that either selects the antenna that can avoid outage or tells the transmitter to train the next $K$ antennas if no such antenna exists. The process continues in the same manner until an antenna that avoids outage is found. If all antennas result in an outage, then the transmitter can simply transmit over an arbitrary antenna. We call this Scheme $\mathtt{B}'$. For the special case of $K=1$, Scheme $\mathtt{B}'$ is exactly the same as the interleaved antenna selection $\mathtt{B}$ in Section \ref{anewantselsch}.

Now, suppose that each training/feedback stage costs $\epsilon$-fraction of the channel codeword time. There are totally $\frac{t}{K}$ stages in Scheme $\mathtt{B}'$ so that the channel capacity is $(1-\frac{t}{K}\epsilon)^+ \log_2(1+ |\langle \mathtt{B}'(\mathbf{h}), \mathbf{h} \rangle|^2 P)$. Given the target data transmission $\rho = \log_2(1+\alpha P)$ as before, the outage probability is given by $\mathrm{Prob}\left( |\langle \mathtt{B}'(\mathbf{h}), \mathbf{h} \rangle|^2 \leq \beta \right)$, where $\beta \triangleq \frac{1}{P}\left((1+\alpha P)^{\frac{1}{(1-\frac{t}{K}\epsilon)^+}} - 1\right)$ can be considered to be a ``modified outage threshold'' that takes into account cost effects of the training/feedback stages. By the definition of Scheme $\mathtt{B}'$, it follows that an outage occurs if and only if $|h_i|^2 \leq \beta,\,\forall i$, and therefore, we have $\mathtt{out}(\mathtt{B}') = (1-e^{-\beta})^t$. After some straightforward calculations, we can also obtain the training length and the feedback rate of the scheme
\begin{align*}
\mathtt{tl}(\mathtt{B}') & = K \frac{1 - (1-e^{-\beta})^t}{1 - (1-e^{-\beta})^K},  & \mathtt{fr}(\mathtt{B}') = \lceil \log_2(1+K) \rceil \frac{1 - (1-e^{-\beta})^t}{1 - (1-e^{-\beta})^K}.
\end{align*}
in closed form. For the special case of $K=1$, and $\beta$ replaced by $\alpha$, the formulae boil down to the ones provided in Section \ref{anewantselsch}. Formally analyzing the tradeoffs between $\mathtt{out}(\mathtt{B}'), \mathtt{tl}(\mathtt{B}')$, and $\mathtt{fr}(\mathtt{B}')$ for given $K$ and $\epsilon$ is not a straightforward task due to the complicated algebraic nature of  expressions. Numerical results in the next section, however, suggest that training antennas one by one is not an optimal strategy in general, and there is an optimal number of antenna groupings $K$ that should be considered.

%

\section{Simulation Results}
\label{secSimulationResults}


\begin{figure}
	\centering
	\includegraphics[width=3.75in]{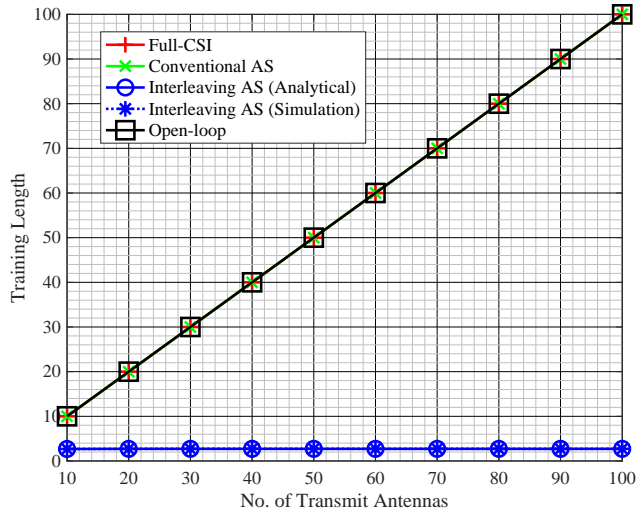}\vspace{-10pt}
	\caption{Training length as a function of $t$ for different schemes in Section \ref{secInterlelele}.}
	\label{trainrate_vs_t}\vspace{-15pt}
\end{figure}

\begin{figure}
	\centering
	\includegraphics[width=3.75in]{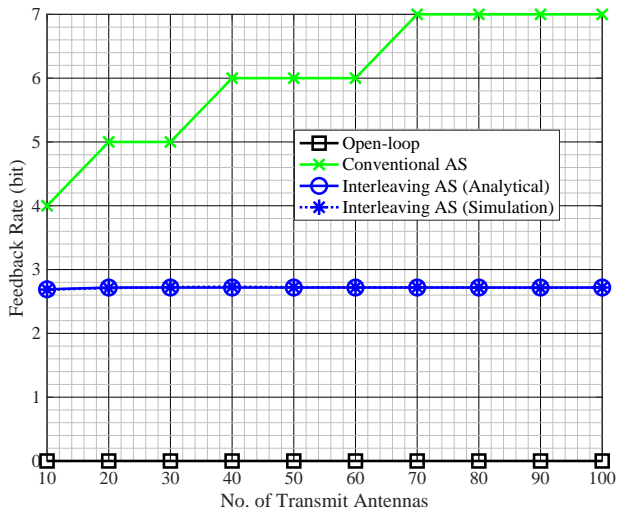}\vspace{-10pt}
	\caption{Feedback rate as a function of  $t$ for different schemes in Section \ref{secInterlelele}.}
	\label{feedbackrate_vs_t}\vspace{-15pt}
\end{figure}

In this section, we provide simulation results to compare the performance of different schemes and quantizers.
Using rate-allocation results in variable rates for different components of the beamforming vector.
We use a Huffman code to send the length of each beamforming vector component.
In other words, each resolution, $\ell_{ij}$, is Huffman coded and the corresponding prefix-free binary codeword representation is sent to the transmitter.
In addition, $\Re x_i$ is quantized by  $q(\Re x_i;\ell_{i1})$ and $\Im x_i$ is quantized by  $q(\Im x_i;\ell_{i2})$, as explained in Section \ref{secVariableRateQuantizer}.

We first present the numerical simulation results of training length and feedback rate as functions of the number of transmit antennas $t$ for different schemes in Section \ref{secInterlelele} in Figs. \ref{trainrate_vs_t} and \ref{feedbackrate_vs_t}, respectively. We abbreviate antenna selection by AS in both figures. In our simulations, we set $\alpha = 1$. Fig. \ref{trainrate_vs_t} shows that as $t$ increases, the average training length of the interleaving antenna selection scheme in Section \ref{anewantselsch} saturates and is lower than those of the full-CSI system, the open-loop system, and the conventional antenna selection scheme in Section \ref{convantenna}. The full-CSI system, the open-loop system, and the conventional antenna selection scheme need to estimate all $t$ channels. Fig. \ref{feedbackrate_vs_t} reveals that as $t$ increases, the average feedback rate of the interleaving antenna selection scheme saturates and is lower than those of the full-CSI system and the antenna selection scheme. Note that the feedback rate of the full-CSI system is infinite. For the interleaving antenna selection scheme in both figures, the simulation results align well with the analytical results provided in Theorem \ref{Thm1}.


\begin{figure}
	\centering
	\includegraphics[width=3.75in]{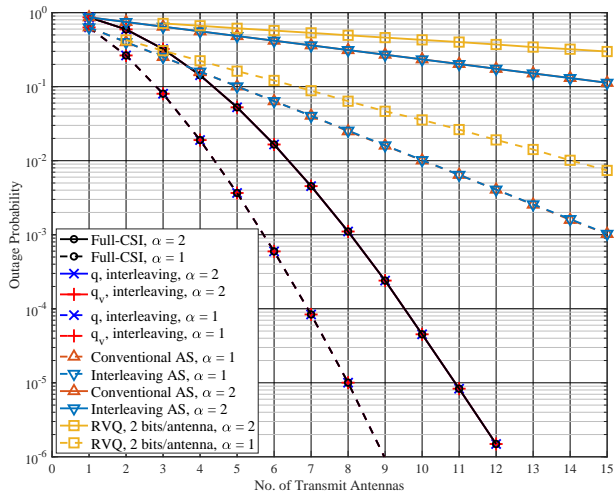}\vspace{-10pt}
	\caption{Outage probability as a function of $t$  for fixed-length and variable-length deadzone quantizers.}
	\label{comparison_outage}\vspace{-15pt}
\end{figure}

\begin{figure}
	\centering
	\includegraphics[width=3.75in]{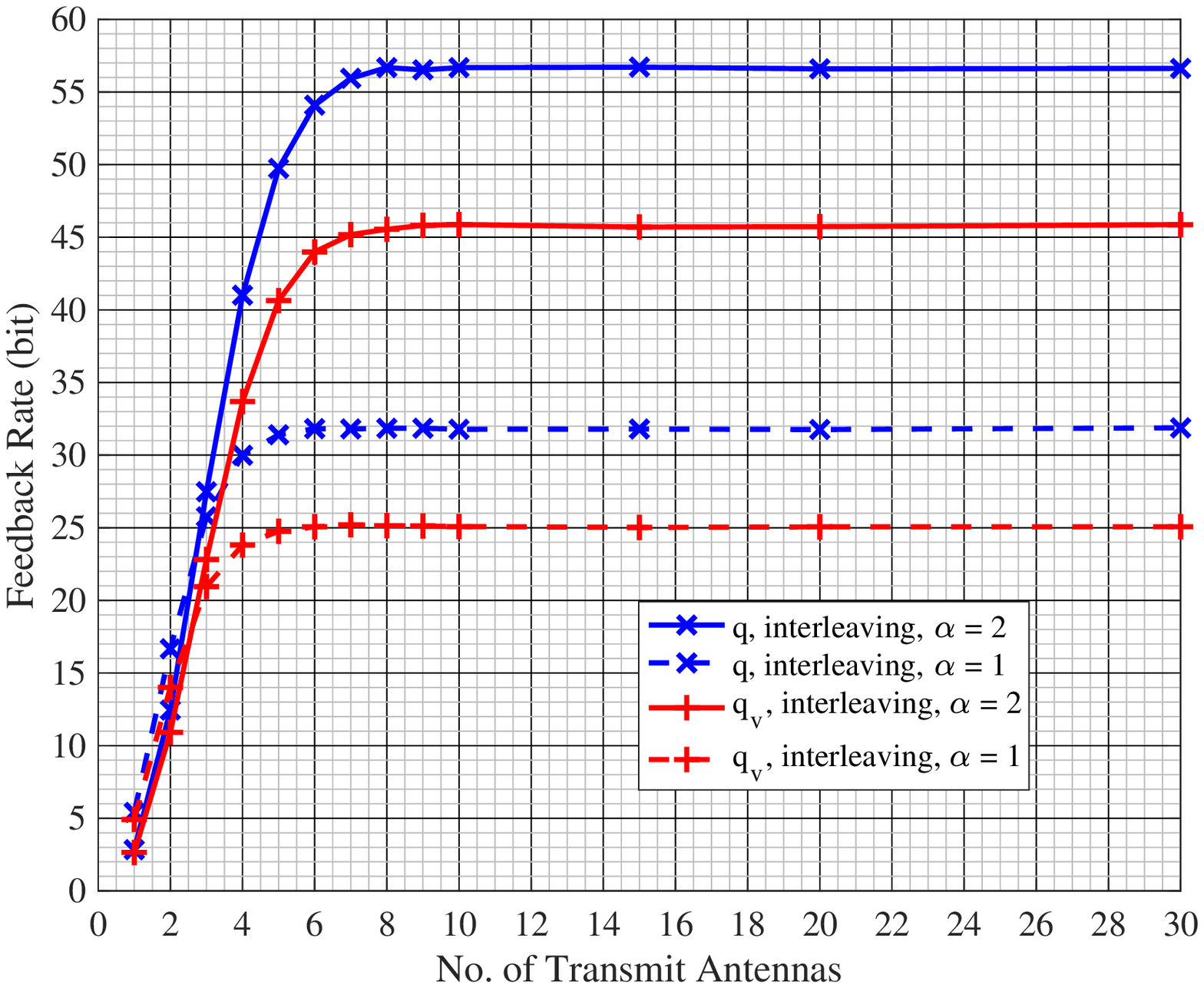}\vspace{-10pt}
	\caption{Feedback rate as a function of $t$ for fixed-length and variable-length deadzone quantizers.}
	\label{comparison}\vspace{-15pt}
\end{figure}

\begin{figure}
	\centering
	\includegraphics[width=3.75in]{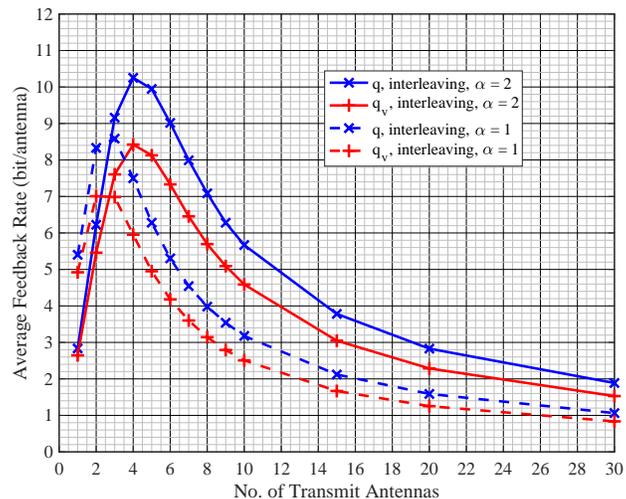}\vspace{-10pt}
	\caption{Average feedback rate as a function of $t$ for fixed-length and variable-length deadzone quantizers.}
	\label{comparison_avgFR}\vspace{-15pt}
\end{figure}

We provide simulation results of the outage probability, the feedback rate, and the average feedback rate as functions of $t$ in Figs. \ref{comparison_outage}, \ref{comparison}, and \ref{comparison_avgFR}, respectively. We consider the deadzone quantizer $q(\vech; \ell)$ and the deadzone quantizer with rate-allocation $q_v(\vech; \boldsymbol\ell)$. The average feedback rate is calculated as the feedback rate divided by the number of transmit antennas. We set $\Delta = 1$ in the rate-allocation algorithm. Fig. \ref{comparison_outage} demonstrates that the interleaving scheme for both quantizers can achieve the same outage probability as the full-CSI system. Fig. \ref{comparison_outage} also shows that the outage probability of the interleaving scheme is better than the outage probabilities of the antenna selection schemes, which is further better than the outage probability of random vector quantization \cite{rvqpaper} with 2 quantization bits per antenna. A smaller outage threshold $\alpha$ leads to a lower outage probability. Fig. \ref{comparison} exhibits several important features: First, the feedback rate with interleaving saturates as $t$ increases. Second, the variable-rate deadzone quantizer $q_v$ reduces the total feedback rate compared to the fixed-rate deadzone quantizer $q$. Third, for the interleaving scheme, the feedback rate decreases as $\alpha$ decreases. This is because a lower resolution for the beamforming vector is acceptable if the outage threshold decreases. According to Fig. \ref{comparison_avgFR}, as the number of transmit antennas $t$ increases, the average feedback rate increases when $t$ is small and deceases when $t$ is large. It is shown that the average feedback rates per antenna for both quantizers are approximately equal to or less than $2$ bits/antenna when $t$ is large.

According to Figs. \ref{comparison} and \ref{comparison_avgFR}, the feedback rates of both deadzone quantizers saturate as the number of transmit antennas increases. This is a key difference compared to the conventional CSI quantization techniques for massive MIMO systems. For example, using the method proposed in \cite{junilNTCQ}, the receiver sends back a binary feedback sequence of length $Bt + q$ where $B$ is the number of quantization bits used per transmit antenna and $q$ is a small positive constant, which scales linearly with the number of transmit antennas. As a result, compared to the conventional CSI quantizers, the proposed deadzone quantizers can save a large amount of feedback overhead when the number of transmit antennas is large.

\begin{figure}
	\centering
	\includegraphics[width=3.75in]{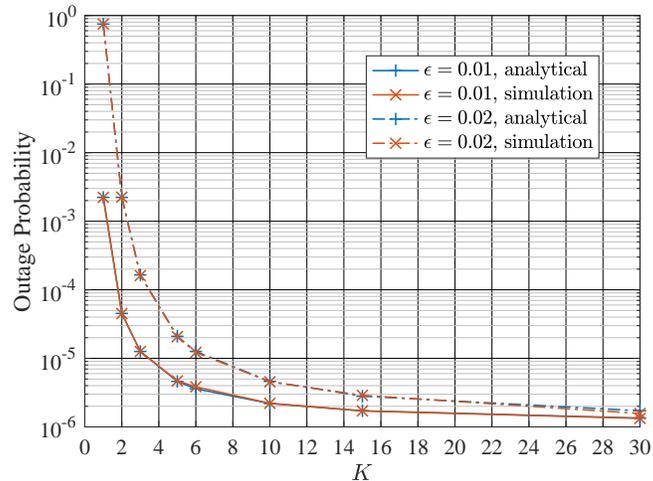}\vspace{-10pt}
	\caption{Outage probability as a function of $K$ for Scheme $\mathtt{B}'$ when $t = 30$, $\alpha = 1$, $P = 1$, $\epsilon = 0.01$ or 0.02.}
	\label{out_eps}\vspace{-15pt}
\end{figure}

\begin{figure}
	\centering
	\includegraphics[width=3.75in]{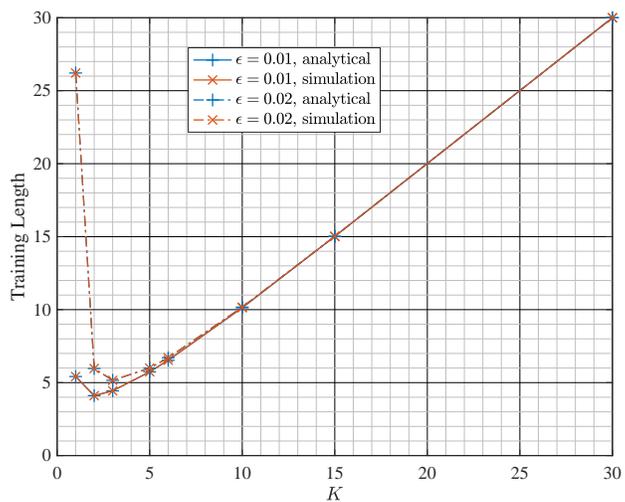}\vspace{-10pt}
	\caption{Training length as a function of $K$ for Scheme $\mathtt{B}'$ when $t = 30$, $\alpha = 1$, $P = 1$, $\epsilon = 0.01$ or 0.02.}
	\label{tr_eps}\vspace{-15pt}
\end{figure}

\begin{figure}
	\centering
	\includegraphics[width=3.75in]{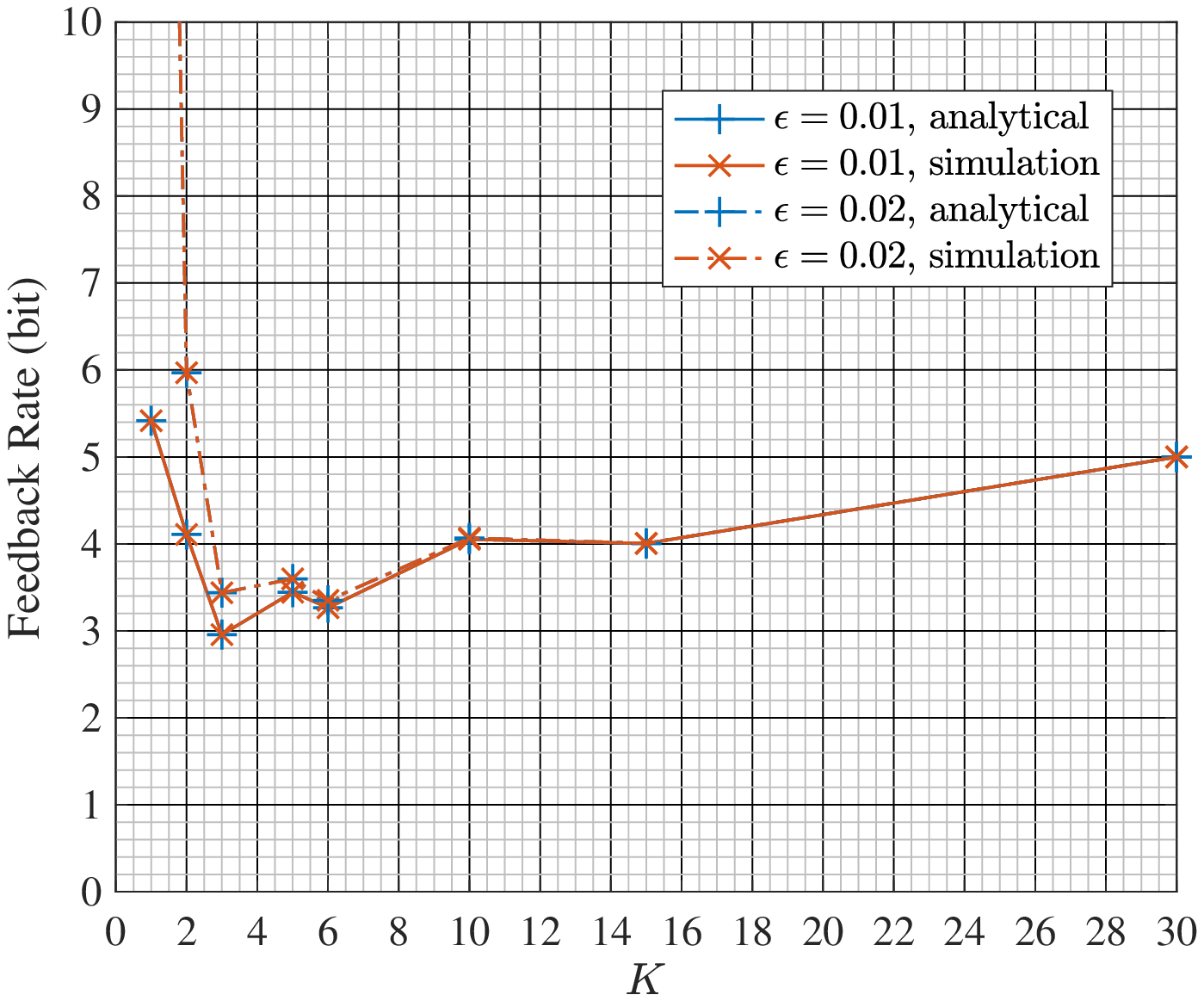}\vspace{-10pt}
	\caption{Feedback rate as a function of $K$ for Scheme $\mathtt{B}'$ when $t = 30$, $\alpha = 1$, $P = 1$, $\epsilon = 0.01$ or 0.02.}
	\label{fr_eps}\vspace{-15pt}
\end{figure}

For Scheme $\mathtt{B}'$ of Section \ref{seclatency}, we present the outage probability, the training length, and the feedback rate as functions of the number of trained antennas at a time, $K$, in Figs. \ref{out_eps}, \ref{tr_eps}, and \ref{fr_eps}, respectively. We can observe that the analytical results match with the simulations in all cases. In Fig. \ref{out_eps}, the outage probability decreases with $K$ since the SNR threshold $\beta$ is a decreasing function of $K$, and $\mathtt{out}(\mathtt{B}')$ decreases as $\beta$ decreases. As expected, as the per-stage cost $\epsilon$ increases, the outage probability increases. Also, according to Fig. \ref{tr_eps}, as $K$ increases from $1$ to $30$, the training length decreases at first but then increases. The optimal value of $K$ that minimizes the training length is $2$ for $\epsilon = 0.01$ and $3$ for $\epsilon = 0.02$. According to Fig. \ref{fr_eps}, the optimal value of $K$ that minimizes the feedback rate is $3$ for $\epsilon = 0.01$ and $6$ for $\epsilon = 0.02$. According to these results, it is suboptimal to train the antennas one by one for the particular choices of the system parameters in Figs. \ref{out_eps}, \ref{tr_eps}, and \ref{fr_eps}. Depending on design requirements, one should consider grouping the antennas in the training and feedback phases.


\section{Conclusion}
\label{secConclusion}
We introduced and analyzed multi-antenna communication schemes whose training and feedback stages are interleaved and mutually interacting. We applied the interleaving scheme to MISO systems to achieve the same outage probability as the full-CSI system using partial CSIT and partial CSIR. We designed a deadzone quantizer and a rate-allocation algorithm to send the feedback messages by a limited number of feedback bits. With $t$ transmit antennas, the interleaving scheme with the deadzone quantizer can achieve a $t$-independent finite feedback rate which only depends on the power constraint and the target data rate. In addition, the rate-allocation algorithm can further reduce the feedback rate by assigning distinct quantization rates to different components in a beamforming vector.

The idea of interleaving can also be used in conjunction with rate adaptation. Suppose the rate-adaptive system can support a number of rates, say, $\rho_1,\ldots,\rho_n$, that one can choose from. Receiver feedback will then be used to choose the beamforming vector as well as the transmission rate. An outage can be declared if the system cannot even support the minimum $\min_{i\in\{1,\ldots,n\}} \rho_i$ of data rates. Given a certain outage probability, one can then study the tradeoff between the feedback rates, training lengths, and the average data transmission rate. For example, supporting high rates typically requires more CSI, and thus larger feedback rates and/or training lengths.

Also, in this work, we have only considered a total power constraint across all antennas. The performance of interleaved training and limited feedback schemes with the additional per-antenna power constraints is another direction for future research. Another interesting topic is the design and analysis of interleaved beam selection schemes for multi-carrier systems such as OFDM.

\appendices
\section{Proof of Proposition \ref{propasymptotes}}
\label{proofofasymptotes}
We first determine the $t\rightarrow\infty$ asymptotic behavior of $\mathtt{out}(\mathtt{F})$. For this purpose, note that
\begin{align}
\label{qwiuepqwiuepew}
\mathtt{out}(\mathtt{F}) = \mathrm{P}(\|\mathbf{h}\|^2 \leq \alpha) = \sum_{i=t}^{\infty} \frac{\alpha^i e^{-\alpha}}{i!},
\end{align}
which leads to an easy lower bound (by considering only the $i=t$ term)
$\mathtt{out}(\mathtt{F}) \geq \frac{\alpha^t e^{-\alpha}}{t!}$. For an upper bound, we can rewrite (\ref{qwiuepqwiuepew}) as
\begin{align}
\label{qiwehqpowepqojj3}
\mathtt{out}(\mathtt{F})
& = \frac{\alpha^te^{-\alpha}}{t!} \sum_{i=0}^{\infty} \frac{\alpha^i}{(t+1)\cdots(t+i)}.
\end{align}
Since $(t+1)\cdots(t+i) \geq i!$, we obtain $\mathrm{P}(\|\mathbf{h}\|^2 \leq \alpha) \leq \frac{\alpha^t}{t!}$. Combining the upper and lower bounds, we have $\mathtt{out}(\mathtt{F})\in \Theta(\frac{\alpha^t}{t!})$, as desired.

We now determine the outage probability of an open-loop system as $t\rightarrow\infty$. We recall that $\mathtt{out}(\mathtt{G}) = \mathrm{P}(\|\mathbf{h}_{\kappa(t)}\|^2 \leq \kappa(t)\alpha)$, where $\kappa(t) \triangleq \arg\min_{k\in\{1,\ldots,t\}}  \mathrm{P}(\|\mathbf{h}_k\|^2 \leq k\alpha)$ with ties broken in favor of $k$ with the smallest index. Then, either $\kappa(t) = t$ for infinitely many $t$ or $\exists t_0 \geq 1,\,\forall t\geq t_0,\,\kappa(t) = t_0$. For values of $\alpha$ that satisfy the latter scenario, we have $\mathtt{out}(\mathtt{G}) = \Theta(1)$.

Suppose $0 < \alpha < 1$. It follows from (\ref{qwiuepqwiuepew}) that $\mathrm{P}(\|\mathbf{h}\|^2 \leq t\alpha) \geq \frac{(t\alpha)^t e^{-t\alpha}}{t!}$. On the other hand, substituting $t\alpha$ instead of $\alpha$ to the expansion in (\ref{qiwehqpowepqojj3}), and using the bound $(t+1)\cdots(t+i) \geq t^i$ for the denominator of the fraction in summation, we obtain $\mathrm{P}(\|\mathbf{h}\|^2 \leq t\alpha) \leq \frac{(t\alpha)^te^{-t\alpha}}{t!(1-\alpha)}$. Combining the upper and lower bounds, it follows that we have $\mathtt{out}(\mathtt{G}) \!=\! \Theta(\frac{(t\alpha)^te^{-t\alpha}}{t!})$ for $0 < \alpha < 1$.

Now, suppose $\alpha \geq 1$. In this case, the Berry-Esseen theorem provides the estimate
$|P(\|\mathbf{h}\|^2 \leq t\alpha) - \Phi((\alpha-1)\sqrt{t})| \leq \frac{C}{\sqrt{t}}$ for some constant $C>0$, where $\Phi(\cdot)$ is the cumulative distribution function of the normal distribution with mean $0$ and variance $1$. It follows that $P(\|\mathbf{h}\|^2 \leq t) \rightarrow \frac{1}{2}$ when $\alpha = 1$, and $P(\|\mathbf{h}\|^2 \leq t\alpha) \rightarrow 1$ whenever $\alpha > 1$. Hence, $\mathtt{out}(\mathtt{G}) = \Theta(1)$ for $\alpha \geq 1$.

\section{Proof of Theorem 2}
\label{proofofschbftheo}
The fact that $\mathtt{out}(\mathtt{D}) = \mathtt{out}(\mathtt{F})$ follows immediately. We thus first calculate the training length $\mathtt{tl}(\mathtt{D})$ of Scheme $\mathtt{D}$. Let $\mathcal{A}_1 \triangleq \{\mathbf{h}\in\mathbb{C}^t:|h_1|^2>\alpha\}$, $\mathcal{A}_i \triangleq \{\mathbf{h}\in\mathbb{C}^t:\|\mathbf{h}_i\|^2 > \alpha,\,\|\mathbf{h}_{i-1}\|^2 \leq \alpha\},\,i=2,\ldots,t$, and $\mathcal{B} \triangleq \{\mathbf{h}\in\mathbb{C}^t:\|\mathbf{h}_t\|^2 \leq \alpha\}$. Note that the sets $\mathcal{A}_1,\ldots,\mathcal{A}_{t},\mathcal{B}$ form a partition of $\mathbb{C}^t$. For any $i\in\{1,\ldots,t\}$, if $\mathbf{h}\in\mathcal{A}_i$, the transmitter trains only the first $i$ channels $\mathbf{h}_1,\ldots,\mathbf{h}_{i}$. If $\mathbf{h}\in\mathcal{B}$, the transmitter trains all the $t$ channels. The training length is thus
\begin{align}
\label{trexpression}
\mathtt{tl}(\mathtt{D}) = \sum_{i=1}^t i \mathrm{P}(\mathbf{h}\in\mathcal{A}_i) +t\mathrm{P}(\mathbf{h}\in\mathcal{B}),
\end{align}
We have $\mathrm{P}(\mathbf{h}\in\mathcal{A}_1) = e^{-\alpha}$. For $i\in\{2,\ldots,t-1\}$, we have
\begin{align}
\label{peycaieval}
\mathrm{P}(\mathbf{h}\in\mathcal{A}_i)   = \int_0^{\alpha} \int_{\alpha-x}^{\infty} e^{-y} \frac{e^{-x}x^{i-2}}{(i - 2)!}\mathrm{d}y\mathrm{d}x =  e^{-\alpha} \int_0^{\alpha}  \frac{x^{i-2}}{(i - 2)!}\mathrm{d}x
 = \frac{\alpha^{i-1}e^{-\alpha}}{(i-1)!}.
\end{align}
Also, since $t\mathrm{P}(\mathbf{h}\in\mathcal{B}) =  \sum_{i=t}^{\infty} t\frac{\alpha^{i} e^{-\alpha} }{i !} \leq \sum_{i=t}^{\infty} \frac{\alpha^{i} e^{-\alpha}}{(i-1)!} = \alpha \sum_{i=t-1}^{\infty}\! \frac{\alpha^ie^{-\alpha}}{i!}$,  we have
\begin{multline*}
\mathtt{tl}(\mathtt{D})  \leq e^{-\alpha}\left(\sum_{i=1}^{t} \underbrace{i}_{=(i-1)+1} \frac{\alpha^{i-1}}{(i-1)!} + \alpha \sum_{i=t-1}^{\infty} \frac{\alpha^i}{i!}\right) \\
 = e^{-\alpha}\left(\sum_{i=2}^{t} \frac{\alpha^{i-1}}{(i-2)!} + \sum_{i=1}^{t} \frac{\alpha^{i-1}}{(i-1)!}  + \alpha \sum_{i=t-1}^{\infty} \frac{\alpha^i}{i!}\right)
 = e^{-\alpha}\biggl(\alpha \underbrace{\sum_{i=0}^{\infty} \frac{\alpha^i}{i!}}_{=e^{\alpha}} + \underbrace{\sum_{i=1}^{t} \frac{\alpha^{i-1}}{(i-1)!}}_{\leq e^{\alpha}}  \biggr)   \leq 1+\alpha,
\end{multline*}
as claimed in the statement of the theorem.

We now calculate the feedback rate $\mathtt{fr}(\mathtt{D})$ of Scheme $\mathtt{D}$. Note that for any $i=1,\ldots,t$, if $\mathbf{h}\in\mathcal{A}_i$, the receiver sends a total of $(i-1)$ bits for requesting the transmitter to train the first $i-1$ antennas (via $i-1$ one-bit binary codewords ``$\mathtt{0}$''). In addition, it sends $2i(L(\mathbf{h}_i)+3)$ bits for the outage avoiding quantized beamforming vector, for a total of $2iL(\mathbf{h}_i)+7i-1$ feedback bits. For $\mathbf{h}\in\mathcal{B}$, there are only $t$ feedback bits. The feedback rate is thus given by
\begin{align*}
& \mathtt{fr}(\mathtt{D}) = \sum_{i=1}^t \int_{\mathcal{A}_i} (2iL(\mathbf{h}_i)+7i-1)f(\mathbf{h})\mathrm{d}\mathbf{h} + t\mathrm{P}(\mathbf{h}\in\mathcal{B}) \\ & = 2\sum_{i=1}^t\! \int_{\mathcal{A}_i}\!\!\! iL(\mathbf{h}_i)f(\mathbf{h})\mathrm{d}\mathbf{h} \!+\! \sum_{i=1}^t (7i\!-\!1)\mathrm{P}(\mathbf{h}\!\in\!\mathcal{A}_i) \!+\! t\mathrm{P}(\mathbf{h}\!\in\!\mathcal{B}),
\end{align*}
where $f(\mathbf{h})$ represents the probability density function of $\mathbf{h}$. According to (\ref{trexpression}), the sum of the last two terms can be upper bounded by $7\mathtt{tl}(\mathtt{D}) \leq 7(1+\alpha)$. Therefore,
\begin{align*}
\mathtt{fr}(\mathtt{D})  \leq 7(1+\alpha) + 2\sum_{i=1}^t \int_{\mathcal{A}_i} iL(\mathbf{h}_i)f(\mathbf{h})\mathrm{d}\mathbf{h}.
\end{align*}
We now evaluate the sum. For this purpose, we partition $\mathcal{A}_1,\ldots,\mathcal{A}_t$ via $\mathcal{A}_i' \triangleq  \{\mathbf{h}\in\mathbb{C}^t:\alpha < \|\mathbf{h}_i\|^2 < 2\alpha,\,\|\mathbf{h}_{i-1}\|^2 \leq \alpha\}$ and $\mathcal{A}_i'' \triangleq \{\mathbf{h}\in\mathbb{C}^t:\|\mathbf{h}_i\|^2 \geq 2\alpha,\,\|\mathbf{h}_{i-1}\|^2 \leq \alpha\}$, with the convention that $\mathbf{h}_0 = 0$ is deterministic. Note that for any $i\in\{1,\ldots,t\}$, if $\mathbf{h}\in\mathcal{A}_i'$, then
\begin{align}
L(\mathbf{h}_i) = \left\lceil \log_2\frac{4i\alpha}{\|\mathbf{h}_i\|^2-\alpha} \right\rceil &\leq 1 + \log_2\frac{4i\alpha}{\|\mathbf{h}_i\|^2-\alpha} \notag \\ &= 3 + \underbrace{\frac{1}{\log 2}}_{\leq 2} \log\frac{i\alpha}{\|\mathbf{h}_i\|^2-\alpha} \leq 3+2\log i+2\log\frac{\alpha}{\|\mathbf{h}_i\|^2-\alpha},
\end{align}
while if $\mathbf{h}\in\mathcal{A}_i''$, then $L(\mathbf{h}_i) = \lceil \log_2 (4i) \rceil \leq 3+2\log i$. Thus,
\begin{multline}
\label{pqihpqiwennenene}
\!\!\!\!\!\mathtt{fr}(\mathtt{D}) \! \leq \! 7(1\!+\!\alpha)\! +\! 6\underbrace{\sum_{i=1}^t i\mathrm{P}(\mathbf{h}\!\in\!\mathcal{A}_i)}_{\triangleq S_1}\! + 4\underbrace{\sum_{i=1}^t i\log i\mathrm{P}(\mathbf{h}\!\in\!\mathcal{A}_i)}_{\triangleq S_2}\! + 4 \sum_{i=1}^t \underbrace{\int_{\mathcal{A}_i'}\!\! i\log\frac{\alpha}{\|\mathbf{h}_i\|^2\!-\!\alpha}f(\mathbf{h})\mathrm{d}\mathbf{h}}_{\triangleq S_{3i}}\!.\!\!\!
\end{multline}
We now find upper bounds on $S_1,S_2$, and $\sum_{i=1}^t S_{3i}$. Regarding $S_1$ and $S_2$, note that we have already evaluated the probabilities $\mathrm{P}(\mathbf{h}\in\mathcal{A}_i),\,i=1,\ldots,t$ in (\ref{peycaieval}). Hence,
\begin{align}
\label{s1bnd}
S_1 = \underbrace{e^{-\alpha}}_{\leq 1} + \alpha e^{-\alpha}\underbrace{\sum_{i=2}^t}_{\leq \sum_{i=2}^{\infty}} \underbrace{\frac{i}{i-1}}_{\leq 2}\frac{\alpha^{i-2}}{(i-2)!} \leq 1+2\alpha,
\end{align}{}\vspace{-20pt}
\begin{multline}
 S_2  = e^{-\alpha} \sum_{i=2}^{t} i \log i \frac{\alpha^{i-1}}{(i-1)!} = e^{-\alpha} \underbrace{\sum_{i=2}^{t}}_{\leq \sum_{i=2}^{\infty}} \underbrace{ \frac{i}{i-1}}_{\leq 2} \log i \frac{\alpha^{i-1}}{(i-2)!}  \leq  2\alpha e^{-\alpha} \sum_{i=0}^{\infty}  \log(i+2) \frac{\alpha^{i}}{i!} \\ =  \! \frac{2\alpha}{e^{\alpha}} \biggl( \sum_{i=0}^{\lceil \alpha \rceil} \underbrace{ \log(i+2)}_{\leq \log(\lceil \alpha \rceil+ 2)}\frac{\alpha^{i}}{i!} + \!\!\! \sum_{i=\lceil \alpha \rceil+1}^{\infty}  \underbrace{\frac{\log(i+2)}{i}}_{\leq \frac{\log( \lceil \alpha \rceil + 2)}{\lceil  \alpha \rceil}}\frac{\alpha^{i}}{(i-1)!}\biggr) \\ \leq  2\alpha e^{-\alpha}\underbrace{\log(\lceil \alpha \rceil+ 2)}_{\leq \log(\alpha+3)}\biggl(\sum_{i=0}^{\lceil \alpha \rceil} \frac{\alpha^{i}}{i!} +  \underbrace{\frac{\alpha}{\lceil  \alpha \rceil}}_{\leq 1} \sum_{i=\lceil \alpha \rceil+1}^{\infty}\frac{\alpha^{i-1}}{(i-1)!}\biggr) \\ \leq 2\alpha e^{-\alpha}\log(\alpha+3)\biggl(\underbrace{\sum_{i=0}^{\lceil \alpha \rceil} \frac{\alpha^{i}}{i!}}_{\leq e^{\alpha}} +  \underbrace{\sum_{i=\lceil \alpha \rceil+1}^{\infty}\frac{\alpha^{i-1}}{(i-1)!}}_{\leq e^{\alpha}}\biggr)
\label{s2bnd}  \leq 4\alpha\log(\alpha+3).
\end{multline}
For an upper bound on $\sum_{i=1}^t S_{3i}$, we consider $S_{31}, S_{32}$, and $\sum_{i=3}^t S_{3i}$ separately. We have
\begin{align}
\label{s31bnd}
S_{31}\! & = \! \int_{\alpha}^{2\alpha}\! \log \frac{\alpha}{x-\alpha}\underbrace{e^{-x}}_{\leq 1} \mathrm{d}x \leq \int_{\alpha}^{2\alpha} \log \frac{\alpha}{x-\alpha} \mathrm{d}x \!=\! \alpha, \\
 S_{32} & =  2\int_0^{\alpha}\!\int_{\alpha - x}^{2\alpha-x} \log\frac{\alpha}{x+y-\alpha}\underbrace{e^{-y}e^{-x}}_{\leq 1}\mathrm{d}y\mathrm{d}x \label{s32bnd} \leq 2\int_0^{\alpha}\!\int_{\alpha - x}^{2\alpha-x} \log\frac{\alpha}{x+y-\alpha}\mathrm{d}y\mathrm{d}x \!=\! 2\alpha^2,
\end{align}

{}\vspace{-20pt}

\begin{multline}
 \sum_{i=3}^tS_{3i}
 = \sum_{i=3}^{t} i \int_0^{\alpha}\int_{\alpha - x}^{2\alpha-x} \log\frac{\alpha}{x+y-\alpha}e^{-y}\frac{x^{i-2}e^{-x}}{(i-2)!}\mathrm{d}y\mathrm{d}x \\
 = \int_0^{\alpha}x\int_{\alpha - x}^{2\alpha-x} \log\frac{\alpha}{x+y-\alpha}e^{-y}\!\!\!\underbrace{\sum_{i=3}^{t}}_{\leq \sum_{i=3}^{\infty}} \!\!\underbrace{\frac{i}{i-2}}_{\leq 3} \frac{x^{i-3}e^{-x}}{(i-3)!}\mathrm{d}y\mathrm{d}x \\
 \leq 3  \int_0^{\alpha}x\int_{\alpha - x}^{2\alpha-x} \log\frac{\alpha}{x+y-\alpha}\underbrace{e^{-y}}_{\leq 1} \underbrace{\sum_{i=3}^{\infty} \frac{x^{i-3}e^{-x}}{(i-3)!}}_{ = 1} \mathrm{d}y\mathrm{d}x \\
 \leq 3  \int_0^{\alpha}x\int_{\alpha - x}^{2\alpha-x} \log\frac{\alpha}{x+y-\alpha}\mathrm{d}y\mathrm{d}x
 \label{s3sumbnd}  = 3\alpha  \int_0^{\alpha}x\mathrm{d}x = \frac{3}{2}\alpha^3.
 \end{multline}
Substituting the bounds in (\ref{s1bnd}), (\ref{s2bnd}), (\ref{s31bnd}), (\ref{s32bnd}), and (\ref{s3sumbnd}) to (\ref{pqihpqiwennenene}), we obtain $\mathtt{fr}(\mathtt{D}) \leq 13+23\alpha + 16\alpha\log(3+\alpha) + 8\alpha^2+6\alpha^3$. Using the bound $\log(3+\alpha) \leq \alpha + 2$, we obtain $\mathtt{fr}(\mathtt{D}) \leq 13+55\alpha+ 24\alpha^2+6\alpha^3$. Finally, the inequalities $\alpha^2 \leq 1+\alpha^3$ and $\alpha \leq 1+\alpha^3$ lead to $\mathtt{fr}(\mathtt{D}) \leq 92+85\alpha^3 \leq 92(1+\alpha^3)$, and this concludes the proof.
\vspace{-10pt}

\linespread{1.5}

\end{document}